\documentclass[aps,pra,twocolumn,preprintnumbers,showpacs,superscriptaddress,amsmath,amssymb]{revtex4}
\usepackage{dcolumn}
\usepackage{bm}
\usepackage{amssymb}
\usepackage[german, english]{babel}
\usepackage{graphicx}
\usepackage{color}
\usepackage{booktabs}
\usepackage{dcolumn}
\usepackage[letterpaper,total={7in,9.5in},top=0.75in,left=0.75in]{geometry}

\begin{document}

\title{Production of quantum degenerate strontium gases:\\Larger, better, faster, colder}

\author{Simon Stellmer}
\affiliation{Institut f\"ur Quantenoptik und Quanteninformation (IQOQI),
\"Osterreichische Akademie der Wissenschaften, 6020 Innsbruck,
Austria}
\author{Rudolf Grimm}
\affiliation{Institut f\"ur Quantenoptik und Quanteninformation (IQOQI),
\"Osterreichische Akademie der Wissenschaften, 6020 Innsbruck,
Austria}
\affiliation{Institut f\"ur Experimentalphysik und
Zentrum f\"ur Quantenphysik, Universit\"at Innsbruck,
6020 Innsbruck, Austria}
\author{Florian Schreck}
\affiliation{Institut f\"ur Quantenoptik und Quanteninformation (IQOQI),
\"Osterreichische Akademie der Wissenschaften, 6020 Innsbruck,
Austria}

\date{\today}




\begin{abstract}
We report on an improved scheme to generate Bose-Einstein condensates (BECs) and degenerate Fermi gases of strontium. This scheme allows us to create quantum gases with higher atom number, a shorter time of the experimental cycle, or deeper quantum degeneracy than before. We create a BEC of $^{84}$Sr exceeding $10^7$ atoms, which is a 30-fold improvement over previously reported experiments. We increase the atom number of $^{86}$Sr BECs to $2.5\times10^4$ (a fivefold improvement), and refine the generation of attractively interacting $^{88}$Sr BECs. We present a scheme to generate $^{84}$Sr BECs with a cycle time of 2\,s, which, to the best of our knowledge, is the shortest cycle time of BEC experiments ever reported. We create deeply-degenerate $^{87}$Sr Fermi gases with $T/T_F$ as low as 0.10(1), where the number of populated nuclear spin states can be set to any value between one and ten. Furthermore, we report on a total of five different double-degenerate Bose-Bose and Bose-Fermi mixtures. These studies prepare an excellent starting point for applications of strontium quantum gases anticipated in the near future.
\end{abstract}

\maketitle

\section{Introduction}
\label{sec:introduction}

Quantum degenerate gases of alkaline-earth-metal atoms, such as strontium, or of atoms with similar electronic structure, such as ytterbium, are attracting considerable interest, fueled by the rich internal structure of these atoms. The two valence electrons give rise to singlet and triplet states, including metastable states, and narrow intercombination lines. In contrast to the bosonic isotopes, the fermionic isotopes possess a nuclear spin, which is crucial for many applications. Proposals demanding some or all of these properties describe the creation of artificial gauge fields \cite{Dalibard2011agp,Gerbier2010gff,Cooper2011ofl,Beri2011zti,Gorecka2011smf}, the implementation of sub-wavelength optical lattices \cite{Yi2008sda}, the processing of quantum information \cite{Stock2008eog}, or the study of many-body systems with dipolar or quadrupolar interaction \cite{Olmos2012lri,Bhongale2012qpo}. The nuclear spin of the fermionic isotopes is at the heart of proposals to study SU($N$) magnetism \cite{Wu2003ess,Wu2006hsa,Cazalilla2009ugo,Hermele2009mio,Gorshkov2010tos,Xu2010lim,FossFeig2010ptk,FossFeig2010hfi,Hung2011qmo,Manmana2011smi,Hazzard2012htp,Bonnes2012alo}, to create non-Abelian artificial gauge fields \cite{Dalibard2011agp,Gerbier2010gff}, to simulate lattice gauge theories \cite{Banerjee2012aqs}, or to robustly store quantum information and perform quantum information processing \cite{Hayes2007qlv,Daley2008qcw,Gorshkov2009aem,Daley2011sdl}. Quantum gas mixtures of alkaline-earth-metal atoms with alkali atoms can be used as a basis for the production of ground-state open-shell molecules, such as RbSr \cite{Zuchowski2010urm,Guerout2010gso}, which constitute a platform towards the simulation of lattice-spin models \cite{Micheli2006atf,Brennen2007dsl}. Bi-alkaline-earth-metal molecules, such as Sr$_2$ \cite{Stellmer2012cou,Reinaudi2012opo}, are sensitive and model-independent probes for variations of the electron-to-proton mass ratio \cite{Zelevinsky2008pto,Kotochigova2009pfa}. Aside from degenerate gases, two-valence-electron atoms have been used for optical clocks \cite{Derevianko2011cpo} and other precision experiments, as well as for the production of ultracold plasmas \cite{Killian2007unp}, and Rydberg gases \cite{Millen2010tee}.

The first alkaline-earth-metal-like element to be cooled to quantum degeneracy was ytterbium \cite{Takasu2003ssb,Fukuhara2007bec,Fukuhara2007dfg,Fukuhara2009aof,Sugawa2011bec,Taie2010roa}, followed by calcium \cite{Kraft2009bec} and strontium \cite{Stellmer2009bec,MartinezdeEscobar2009bec}. Strontium has one fermionic and three bosonic stable isotopes, with vastly different natural abundances and scattering lengths, see Tab.~\ref{tab:SrIsotopes}. All isotopes have been cooled to quantum degeneracy using slightly different approaches to accommodate their respective properties. The first isotope brought to quantum degeneracy was the low abundance isotope $^{84}$Sr, since its interaction properties are ideal for evaporative cooling \cite{Stellmer2009bec,MartinezdeEscobar2009bec}. Next, the $^{88}$Sr isotope, which has a very small and negative scattering length, was Bose condensed by sympathetic cooling with $^{87}$Sr \cite{Mickelson2010bec}. The last remaining bosonic isotope, $^{86}$Sr, has a very large scattering length, which required evaporative cooling at relatively low densities to avoid three-body decay \cite{Stellmer2010bec}. The fermionic isotope $^{87}$Sr was cooled to quantum degeneracy in both a mixture of spin states \cite{DeSalvo2010dfg} and as a spin-polarized sample together with $^{84}$Sr \cite{Tey2010ddb}.

\begin{table}[b]
	\centering
\caption{Important properties of the four stable strontium isotopes. The scattering lengths $a$ are averages of values taken from \cite{Martinezdeescobar2008tpp} and \cite{Stein2010tss}. Only the fermionic $^{87}$Sr isotope has a nuclear spin $I$.}
	\label{tab:SrIsotopes}
		\begin{tabular*}{\columnwidth}{@{\extracolsep{\fill}}ccdrc}\hline\hline \noalign{\smallskip}
&\multicolumn{1}{c}{statistics}&\multicolumn{1}{c}{abundance}&\multicolumn{1}{c}{$a$}&\multicolumn{1}{c}{$I$}\\
& &\multicolumn{1}{c}{(\%)}&\multicolumn{1}{c}{($a_0$)}&\\
                      \noalign{\smallskip}\hline\noalign{\smallskip}
			$^{84}$Sr & bosonic   &  0.56 & 123 &   0 \\
			$^{86}$Sr & bosonic   &  9.86 & 800 &   0 \\
			$^{87}$Sr & fermionic &  7.00 &  96 & 9/2 \\
			$^{88}$Sr & bosonic   & 82.58 &  -2 &   0 \\ \hline \hline
		\end{tabular*}
\end{table}

In this article, we describe our scheme to create quantum degenerate samples of strontium, which improves several key properties substantially compared to previously reported results. In particular, we create quantum gases with much larger atom number, or much shorter time of the experimental cycle, which increases signal to noise and data rate in many applications and gives good prospects for using Sr as a coolant for other elements. The temperature of degenerate Fermi gases has been lowered to $T/T_F=0.10(1)$, which is crucial for the exploration of SU($N$) magnetism. Furthermore, we report on the production of novel isotopic quantum gas mixtures. These improvements are obtained with a universal experimental setup that can be used for all isotopes, as well as mixtures among them, with only little modifications.

The article is structured as follows: in Sec.~\ref{sec:ImprovedExperimentalSetup}, we prepare the stage by describing the experimental setup. In Sec.~\ref{sec:84BEC} we present the creation of a large $^{84}$Sr BEC containing more than $10^7$ atoms and in Sec.~\ref{sec:fastBEC} a route to BECs containing $10^5$ atoms with a $2\,$s cycle time. In Sec.~\ref{sec:86BEC} we show improved experiments on BECs of the $^{86}$Sr isotope and in Sec.~\ref{sec:88BEC} a robust route to attain BECs of the $^{88}$Sr isotope. In Sec.~\ref{sec:BoseBose} we introduce two different double-degenerate Bose-Bose mixtures. In Sec.~\ref{sec:87DFG} we turn to the fermionic isotope and present deeply degenerate samples with different numbers of occupied spin states. In Sec.~\ref{sec:BoseFermi} we present mixtures of the fermionic with each of the bosonic isotopes. A summary of all experiments is given in Tab.~\ref{tab:Summary} and a short conclusion follows in Sec.~\ref{sec:conclusion}.

\section{Experimental setup and procedures}
\label{sec:ImprovedExperimentalSetup}

In this section, we will describe crucial parts of the experimental setup and give an account of the experimental sequence. The basic scheme to produce Sr quantum gases is the preparation of a laser cooled sample in a magneto-optical trap (MOT), the transfer of this sample into an optical dipole trap, followed by evaporative cooling. Laser cooling of Sr atoms has already been described in detail elsewhere \cite{Katori1999mot,Mukaiyama2003rll,Sorrentino2006lca,Boyd2007PhD,Stellmer2009bec,Tey2010ddb} and will be reviewed only briefly in Sec.~\ref{sec:ImprovedExperimentalSetupMOT}. In Sec.~\ref{sec:ImprovedExperimentalSetupDipoleTrap} we describe the design of a flexible optical dipole trap, which has proven to be the crucial element for efficient evaporation. The transfer of atoms from the MOT into the dipole trap is described in Sec.~\ref{sec:ImprovedExperimentalSetupDipoleTrapLoading}. Evaporative cooling to quantum degeneracy is discussed in Sec.~\ref{sec:ImprovedExperimentalEvaporativeCooling} and data acquisition and analysis in Sec.~\ref{sec:ImprovedExperimentalDataAnalysis}. A detailed description of the experimental setup and the laser systems can be found in Refs.~\cite{WillePhD,StellmerPhD}.

\begin{figure}[htp]
\includegraphics[width=\columnwidth]{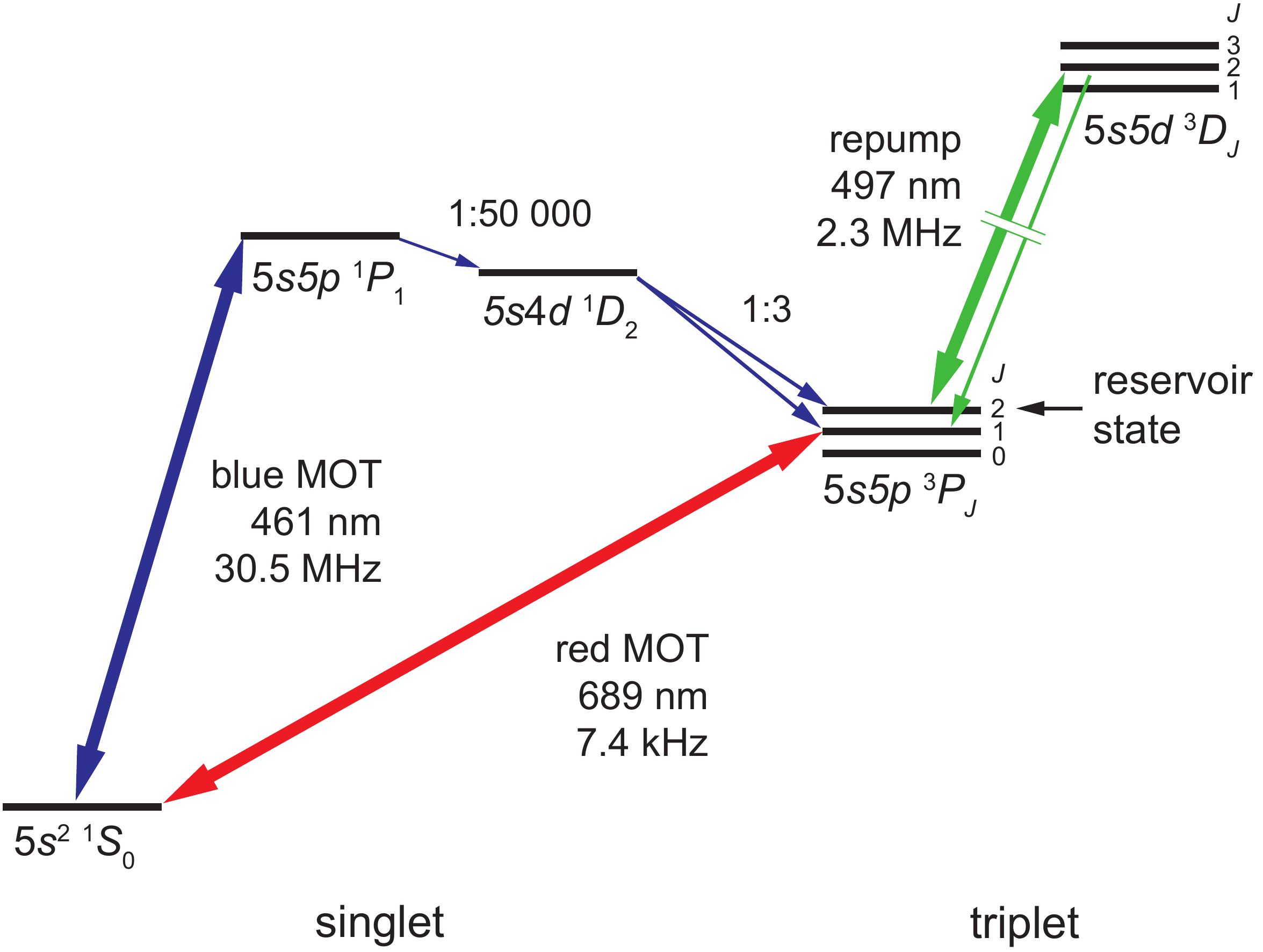}
\caption{\label{fig:Fig1_SrLevelScheme} (Color online) Schematic illustration of the energy levels and transitions used for cooling and trapping of Sr atoms. The blue MOT is operated on the strong $^1S_0$-$^1P_1$ transition. The loading of the magnetic trap proceeds via the weak leak of the excited state (branching ratio 1:50\,000) into the $^1D_2$ state, which itself decays with a 1:3 probability into the metastable $^3P_2$ state. Here the atoms can be magnetically trapped and accumulated for a long time, forming a reservoir. The transition $^3P_2$-$^3D_2$ allows us to depopulate the metastable state by transferring the atoms into the $^3P_1$ state. The latter represents the excited state of the $^1S_0$-$^3P_1$ intercombination line, used for narrow-line cooling in the red MOT.}
\end{figure}

\subsection{Laser cooling}
\label{sec:ImprovedExperimentalSetupMOT}

For the preparation of a laser cooled sample of Sr in an ultrahigh vacuum chamber, we use a variation of standard laser cooling and trapping techniques \cite{Metcalf1999book}. The preparation method has two stages and is based on the term scheme of Sr [see Fig.~\ref{fig:Fig1_SrLevelScheme}]. During the first stage, a transversally cooled and Zeeman-slowed atomic beam is captured in a ``blue'' MOT. The MOT cycle is not completely closed and pumps atoms into a metastable magnetic state, in which they can be accumulated in a magnetic trap. During the second stage, the atoms are pumped out of the metastable state and further cooled in a ``red'' MOT. This MOT uses a narrow-linewidth transition and laser cools the sample to a phase-space density approaching 0.1.

All laser cooling steps of the first stage use the broad $^1S_0 - {^1P_1}$ transition at 461\,nm. The linewidth of this blue transition is $\Gamma_{\rm blue}=2\pi \times 30.5\,$MHz, which corresponds to a comparatively high Doppler temperature of 720\,$\mu$K. The transition is nearly closed and no repumping is needed to operate the blue MOT. The isotope shift of about 70\,MHz per mass unit is larger than the transition linewidth, which results in isotope selective laser cooling. Operation can be switched between isotopes within 100\,ms by changing the laser frequency. Our laser source is a frequency doubled diode laser system with 350\,mW output power.

\begin{figure}[htp]
\includegraphics[width=\columnwidth]{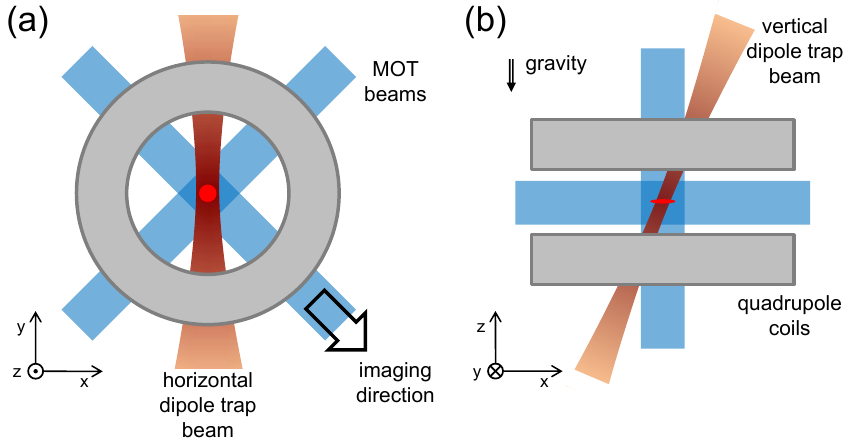}
\caption{\label{fig:Fig2_Setup} (Color online) Schematic view of the setup as seen from above (a) and from the side (b). Shown are the quadrupole field coils (light gray areas with dark gray outline), the MOT beams (rectangular blue areas), the elliptical horizontal and the circular vertical dipole trap beams (bow tie shaped red areas), and the direction of imaging, the ($\hat{\mathbf x}-\hat{\mathbf y}$)-direction (open arrow). The horizontal dipole trap is elongated in the x-direction and quantum degenerate gases are created in the pancake shaped crossing region of the two dipole trap beams. For clarity, only the focus of the horizontal (vertical) dipole trap beam is shown in the side (top) view.}
\end{figure}

During the first stage, Sr atoms are emitted through a bundle of microtubes ($\sim 70$ steel tubes, 200\,$\mu$m inner diameter, 300\,$\mu$m outer diameter, 12\,mm length) from an oven heated to about $600^{\rm{\circ}}$C. The resulting atomic beam is transversally cooled by two retroreflected laser beams, orthogonal to the atomic beam and to each other, which increases the flux of atoms captured in the MOT by about a factor four. The transversal cooling beams have an elliptical beam profile elongated along the direction of the atomic beam (15\,mm by 3\,mm waist), a detuning of $-15\,$MHz from resonance, and a power of 10\,mW per beam. The atomic beam passes two consecutive differential pumping stages, helping to achieve excellent vacuum conditions in the ultrahigh vacuum chamber, which result in a lifetime of two minutes for Sr samples stored in an optical dipole trap. The atoms are decelerated by a spin-flip Zeeman-slower of 80\,cm length. The Zeeman-slower beam has a power of  $\sim 35\,$mW, a detuning of $-430\,$MHz, a waist of about 8\,mm at the position of the MOT, and is slightly focussed onto the aperture of the oven. The atoms are captured by the blue MOT in a glass cell vacuum chamber. The MOT consists of three orthogonal retroreflected laser beams, with a detuning of -30\,MHz [see Fig.~\ref{fig:Fig2_Setup}]. The horizontal (vertical) MOT beams have a waist of 5\,mm (4\,mm) and a power of 4\,mW (1\,mW) each. The quadrupole magnetic field of the MOT has a gradient of 55\,G/cm in the axial direction (z-direction).

The blue MOT cycle is not completely closed, as atoms in the $^1P_1$ state can decay into the $^1D_2$ state and further into the metastable $^3P_2$ state, which is magnetic and has a natural lifetime of many minutes \cite{Yasuda2004lmo}. Weak-magnetic-field seeking atoms in this state are trapped in the magnetic quadrupole field that is also used for the MOT, forming a reservoir of atoms. We operate the MOT until an estimated amount of $10^8$ atoms are captured in this metastable state reservoir, which usually takes between 100\,ms and a few seconds, depending on the natural abundance of the isotope of interest. At this point, all laser beams and the Zeeman slower magnetic field coils are switched off and the atomic beam is blocked by a mechanical shutter. The use of the metastable state reservoir for accumulation of atoms is an elegant way to overcome the low natural abundance of the $^{84}$Sr isotope (0.56\%), or to store a certain isotope while loading a different one. The lifetime of atoms in the reservoir is about 30\,s.

During the second laser cooling stage, the atoms are further cooled in a red MOT operated on the $^1S_0 - {^3P_1}$ intercombination line at a wavelength of 689\,nm. The linewidth of this transition is $\Gamma_{\rm red}=2\pi \times 7.4\,$kHz, which corresponds to a Doppler temperature of only 180\,nK. This temperature is comparable to the recoil temperature, which is 460\,nK. The saturation intensity of this transition is $I_{\rm sat}=3\,\mu{\rm W/cm}^2$. The red MOT beams are superimposed with the blue MOT beams, have a waist of 3\,mm, and an initial peak intensity of $6\,{\rm mW/cm}^2=2000\,I_{\rm sat}$ per beam. Our laser source is based on a extended cavity diode laser with a linewidth of about 2\,kHz and an absolute stability better than 500\,Hz. The narrow linewidth is achieved by locking the laser to a cavity, which is stabilized in length using a spectroscopy lock on a timescale of 5\,s. The required frequency components are derived from this master laser by acousto-optical modulators (AOMs) and amplified by slave diode lasers.

Loading of the red MOT is accomplished by optical pumping on the $^3P_2 - {^3D_2}$ transition at 497\,nm [see Fig.~\ref{fig:Fig1_SrLevelScheme}]. The repump beam has a power of a few mW and a waist of 8\,mm. Its laser source is a frequency doubled diode laser, which is referenced to a stable cavity. After switching the repump beam on, the magnetic field gradient is lowered to 1.15\,G/cm within 0.1\,ms. The temperature of the atoms that are pumped out of the metastable state reservoir is on the order of the blue MOT Doppler temperature, which is 720\,$\mu$K. To obtain a capture velocity that is high enough for these atoms and to increase the capture volume, we frequency modulate the red MOT light. The modulation creates sidebands that cover a detuning range between $-150\,$kHz and $-8\,$MHz with a spacing of 20\,kHz. The red MOT is kept in these conditions for 50\,ms to capture the atoms.

To increase the phase-space density, the red MOT is compressed by ramping the detuning, modulation width, and power of the MOT beams to lower absolute values over the next $\sim 500\,$ms. The modulation is switched off already after 100\,ms of this ramp. At the end of the ramp, the detuning of the MOT beams is typically $-150\,$kHz and the peak intensity is $1.5\,\mu{\rm W/cm}^2=0.5\,I_{\rm sat}$ per beam. This compression phase leads to a colder and denser sample. A few $10^7$ atoms can be cooled to temperatures of about 400\,nK for the almost non-interacting $^{88}$Sr isotope or about 800\,nK for the other isotopes, including the fermionic one. The attainable temperature highly depends on the density of the sample, which creates a trade-off between temperature and atom number.

\begin{figure}[htp]
\includegraphics[width=\columnwidth]{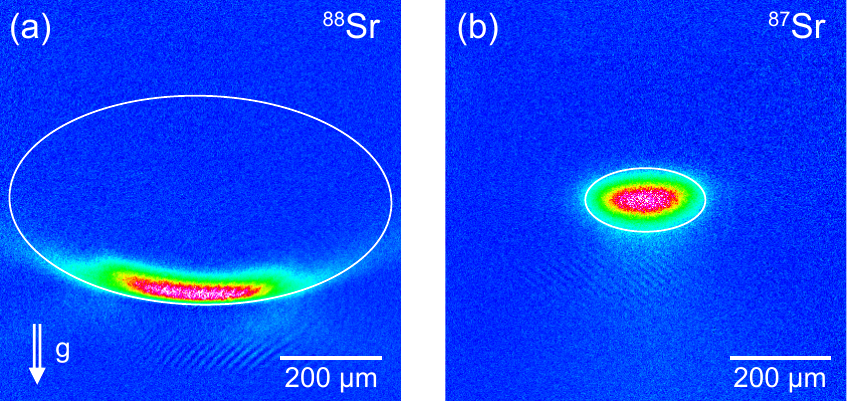}
\caption{(Color online) Narrowline MOTs of (a) the bosonic $^{88}$Sr and (b) the fermionic $^{87}$Sr isotopes, shown by \textit{in-situ} absorption images taken along the horizontal direction. In case (a), the MOT beams have a detuning of -50\,kHz and a peak intensity of $I_{\rm sat}$. Gravity and laser cooling forces balance each other on the surface of an ellipsoid, which has a vertical radius of $200\,\mu$m, giving rise to a pancake shaped MOT. In the fermionic case (b), we operate at a detuning of about -20\,kHz and an intensity equal to $I_{\rm sat}$ for both MOT frequency components. The atoms occupy the volume of an ellipsoid. In both cases, the magnetic field gradient is $\partial B/\partial z=1.15\,$G/cm, the atom number $1.3(1)\times10^{6}$, and the temperature $\sim 700\,$nK. The white ellipses are a guide to the eye.}
\label{fig:Fig3_BoseFermiRedMOT}
\end{figure}

The operation of the fermionic red MOT is slightly complicated in comparison to the bosonic red MOT, because of the existence of hyperfine structure in fermionic $^{87}$Sr, which is absent in bosonic Sr isotopes. Since $^{87}$Sr possesses a nuclear spin of $I=9/2$, its $^3P_1$ state is split into three hyperfine states with $F'=11/2$, 9/2, and 7/2. A long-lived MOT is obtained by using the transitions to the $^3P_1(F'=11/2)$ and the $^3P_1(F'=9/2)$ states simultaneously \cite{Mukaiyama2003rll,Boyd2007PhD}. During the capture and compression phases of the red MOT, both laser components are ramped as described above.

After the compression phase, the shapes of the bosonic and fermionic red MOTs are very different [see Fig.~\ref{fig:Fig3_BoseFermiRedMOT}]. In the bosonic case, the atoms occupy the surface of an ellipsoid, defined by the resonance condition $- h \Delta=g_J \mu_B B(\mathbf{r})$, where $h$ is the Planck constant, $\Delta$ the detuning of the MOT laser from the zero magnetic field resonance, $g_J=1.5$ the Land\'e $g$-factor of the $^3P_1$ state, $\mu_B$ the Bohr magneton,  and $B(\mathbf{r})$ the magnitude of the MOT quadrupole magnetic field at location $\mathbf{r}$. On the lower part of this surface the force of gravity is comparable to the restoring light force, and atoms accumulate there \cite{Katori1999mot,Loftus2004nlc}. In our typical operation conditions ($\Delta=-150\,$kHz, $\partial B/\partial z=1.15\,$G/cm) the atoms form a disc with about $500\,\mu$m diameter, located $600\,\mu$m below the quadrupole center. By contrast, the fermionic MOT occupies the entire volume of an ellipsoid. In order to achieve densities similar to the bosonic case, we reduce the detuning of both MOT frequency components to a few linewitdhs, which reduces the size of the ellipsoid. The positions of the bosonic and fermionic red MOTs are thus different, which needs to be taken into account for dipole trap loading.

Our laser cooling scheme can be easily extended to prepare a mixture of two Sr isotopes. For this task, the first cooling stage is executed twice, filling the long-lived metastable state reservoir consecutively with the first, then with the second isotope. The only difference between these two executions is the wavelength of the blue laser system, which is changed from one isotope to the other. The second laser cooling stage is executed simultaneously for both isotopes. Here, the red MOT beams and the repump laser contain several frequency components each, tuned to the transitions required for repumping and laser cooling of the two isotopes. The required frequency components can be easily derived from master lasers using AOMs.

\subsection{Dipole trap setup}
\label{sec:ImprovedExperimentalSetupDipoleTrap}

For further storage and cooling of the atomic cloud, we use a crossed beam optical dipole trap \cite{Grimm2000odt}, consisting of a horizontal and a (near-)vertical infrared laser beam [see Fig.~\ref{fig:Fig2_Setup}]. The dipole trap has its cross at the location of the atomic cloud and is switched on at the beginning of the red MOT phase. Each laser beam is derived from a separate multimode 5-W Yb-doped fiber laser (IPG YLD-5-1064-LP) operating at 1065\,nm. The intensity of each laser beam is set, but not actively stabilized, by AOMs. Polarization-maintaining single-mode optical fibers behind the AOMs are used for mode-cleaning.

The transfer of atoms from the red MOT into the dipole trap is optimized by spatial mode-matching. Since the bosonic red MOT is pancake-shaped, we give the horizontal dipole trap an elliptical focus, elongated in the horizontal plane. To shape the focus, we use a cylindrical telescope between the output collimator of the mode-cleaning fiber and the lens focussing the beam onto the atoms. The beam has a vertical waist of $18(2)\,\mu$m and a horizontal waist of $250(50)\,\mu$m, thus an aspect ratio of 15. With a typical power of 2\,W used for loading of the dipole trap, the trap frequencies are $f_x = 40(4)$\,Hz, $f_y = 6(1)$\,Hz, and $f_z = 620(50)$\,Hz. The trap is $k_B\times 10(1)\,\mu$K deep, where $k_B$ is the Boltzmann constant and we take into account gravitational sagging.

The vertical dipole trap beam is used to obtain a stronger confinement in the horizontal plane, but it does not support the atoms against gravity. This beam is focused into the vacuum chamber by a spherical lens and is centered onto the focus of the horizontal dipole trap beam. The $1/e^2$-radius of the vertical beam in the plane of the horizontal dipole trap can be set to any value between 20\,$\mu$m and 300\,$\mu$m, and is optimized separately for evaporative cooling of each Sr isotope or isotopic mixture. Two methods are used to adjust the beam size. Coarse tuning is done by changing the size of the collimated beam after the mode-cleaning fiber by exchanging the fiber collimator. Fine tuning is done by moving the position of the focussing lens along the beam axis using a translation stage. Due to optical restrictions, this dipole trap beam is not exactly vertical but propagates at an angle of $22^{\rm{\circ}}$ with respect to gravity. To be able to quickly switch between two different waists, two vertical dipole trap beams are implemented, reaching the atoms under angles that differ by less than 30\,mrad. In all experiments presented here, only either one or the other of the two beams is used.

The role of the horizontal and vertical dipole trap beams for evaporative cooling is clearly distinct. The horizontal dipole trap beam sets the trap depth, since atoms leave the trap predominantly vertically downwards, aided by gravity. It also determines the vertical trap oscillation period and thereby the timescale on which atoms can escape the trap. Evaporative cooling can be limited by this timescale and therefore a high vertical trap frequency is desirable. The vertical dipole trap adds confinement in the horizontal plane and is a means to tune the density of the sample. This additional confinement is used in all experiments presented, but for the creation of a $^{84}$Sr BEC with high atom number. To achieve a sufficiently high density for evaporative cooling, we never need horizontal trap frequencies exceeding the vertical trap frequency. The trap has always a pancake shape, which is fortuitous for two reasons. First, this trap shape is matched to the bosonic red MOT, which is a requirement for efficient trap loading. And second, evaporation happens across a large surface of the trap, which is good for evaporation efficiency.

To determine the potential of the dipole trap, we perform trap oscillation measurements. The measurements are typically performed using a nearly pure BEC in a trap that is slightly recompressed from the conditions in which the BEC was created. To measure the trap oscillation frequencies in the radial plane, we use the additional vertical dipole trap beam to excite sloshing modes in the horizontal plane. To measure the vertical trap oscillation frequency, we exploit gravitational sagging and excite the vertical sloshing mode by a sudden compression of the horizontal dipole trap. From the trap frequencies and the power of the dipole trap lasers, we derive the $1/e^2$-radii of the dipole trap beams at the location of the cross and the depth of the potential, taking into account the effect of gravitational sagging. Knowing the radii of the dipole trap beams, we can calculate the trap potential for any isotope and dipole beam power. For cold atomic clouds near quantum degeneracy, the potential is usually well approximated by a harmonic potential. This approximation is not always valid for atomic clouds with MOT temperatures, since these clouds can be larger in the horizontal direction than the size of the vertical dipole trap beam.

\subsection{Dipole trap loading and performance}
\label{sec:ImprovedExperimentalSetupDipoleTrapLoading}

In the following, we will discuss loading of atoms into the dipole trap and then address two important details. The first detail are the frequency shifts induced by the dipole trap and by the magnetic field inhomogeneity and the second detail are the requirements for reliable dipole trap loading.

To load atoms into the dipole trap, the spatial overlap between the red MOT and the dipole trap is fine-tuned by shifting the MOT quadrupole magnetic field center using an offset magnetic field. About 50\% of the atoms are transferred from the MOT into the dipole trap. After the atoms are in the dipole trap, the MOT light is kept on for another 100\,ms at an intensity of $\sim 0.5\,I_{\rm sat}$ and a detuning of $-20\,$kHz from the Zeeman- and light-shifted cooling transition in the trap center. During this time, the atoms that are initially spread out in the horizontal dipole trap beam are pushed into the center of the trap by the horizontal MOT beams, thereby increasing the density of the sample. Afterwards all near-resonant laser fields are ramped off. Since the location of the bosonic and fermionic red MOTs are different [see Fig.~\ref{fig:Fig3_BoseFermiRedMOT}], preparation of a Bose-Fermi mixture in the dipole trap requires a sequential loading scheme. We first load the fermions and then shift the quadrupole center upwards by applying an offset magnetic field to load the bosons. We can transfer up to $4\times 10^7$ atoms into the dipole trap with a shot-to-shot variation of about 1\%. The temperature of the cloud is between 400\,nK and 1.5\,$\mu$K, depending on isotope and density. Since the ground state is non-magnetic for bosonic Sr, the quadrupole magnetic field used for the MOT does not create a potential on the atoms and is typically left on during evaporation. If fermions are used, the quadrupole field is switched off and a homogeneous magnetic field of 3\,G is applied during evaporation. After loading the dipole trap, the gas is left to thermalize for typically 250\,ms. Remarkably, the phase-space density at this point in the experimental sequence can reach up to 0.3, which shows the power of narrow-line laser cooling. The lifetime of a dilute sample, measured after some evaporative cooling and a slight recompression of the dipole trap, is two minutes, much longer than typical evaporative cooling timescales. The heating rate is below 1\,nK/s.

An important effect influencing in-trap cooling is the differential light shift between the two states connected by the laser cooling transition. The different ac-polarizabilities of the two states together with the position-dependent intensity of the dipole trap beams leads to a position-dependent laser cooling transition frequency across the sample. For optimal laser cooling the differential light shift needs to be minimized. The magnitude of the differential light shift depends on the dipole trap wavelength, the orientation of the magnetic field, and the polarizations of dipole trap and cooling light \cite{Boyd2007PhD}. We minimize the light shift for laser cooling of a bosonic red MOT. In this case, the magnetic field across the atomic cloud is nearly homogeneous and oriented vertically, since the quadrupole field center is located 600\,$\mu$m above the cloud. Only the $\sigma^-$ transition is near resonant with the cooling light. In this simple situation and for our dipole trap wavelength of 1065\,nm, it is possible to reduce the differential light shift to a small value by choosing optimal polarizations for the dipole trap beams. The vertical dipole trap beam has to be polarized to the opposite circular direction of the vertically upwards propagating MOT beam. The horizontal dipole trap beam has to be linearly polarized in the vertical direction.

The differential light shift is determined to within 1\,kHz using absorption imaging on the laser cooling transition \cite{Stellmer2011dam} and found to be positive and below +20\,kHz for all dipole trap settings used. Since this value is larger than the transition linewidth of 7.4\,kHz, the light shift needs to be taken into account when choosing the optimum frequency for laser cooling of a sample confined in the dipole trap. Inconveniently, the optimum laser frequency would depend on the location of an atom in the trap, if the light shift across the sample would be too inhomogeneous. Fortunately the inhomogeneity of the light shift is only $\sim 2\,$kHz, since the temperature of the gas during the dipole trap loading process corresponds to only one tenth of the trap depth. This inhomogeneity is smaller than the transition linewidth and comparable to the linewidth of our cooling laser. Therefore the laser cooling process at different locations in the sample is not significantly changed by the light shift.

An additional frequency shift for atoms in the dipole trap is created by the different absolute magnetic field values at different locations in the trap. For the conditions under which we transfer bosonic atoms from the MOT into the dipole trap, this shift corresponds to 10\,kHz across the sample, which corresponds to 1.5 linewidths and is tolerable.

Reliable transfer of atoms from the bosonic red MOT into the dipole trap requires high relative position stability between these two traps. This requirement is especially strict in the vertical direction, since the horizontal dipole trap beam has a vertical waist of only $18\,\mu$m and the extension of the red MOT in the vertical direction is not much bigger. While the dipole trap position is very stable, the vertical position of the red MOT depends on the cooling laser frequency and on the position of the quadrupole magnetic field center, which both can easily fluctuate. To illustrate the effect, we consider the dependence of the vertical red MOT position in absence of the optical dipole trap on these two parameters. With the vertical magnetic field gradient of 1.15\,G/cm, the red MOT is displaced $18\,\mu$m by a change in the offset magnetic field of only 2\,mG. The same displacement is reached for a frequency change of 4\,kHz (see the resonance condition describing the shape of the bosonic red MOT above).

We now quantify the stability of the vertical offset magnetic field and the laser frequency reached in the experiment. We first measure the root mean square temperature variation of a dipole trapped sample over a few dozen experimental runs and obtain $\sim 10\,$nK. Then we measure the relation of the temperature with the detuning of the cooling light and obtain 25\,nK/kHz over a range of 20\,kHz. Assuming that only the laser frequency fluctuates, we deduce an absolute laser frequency stability of better than 400\,Hz. The same temperature variation can be caused by a vertical offset magnetic field fluctuation of $200\,\mu$G. Because of the light shift induced by the dipole trap on the cooling transition, also a slight change of the polarization of the dipole trap beams could contribute to the observed temperature variation.

\subsection{Evaporative cooling}
\label{sec:ImprovedExperimentalEvaporativeCooling}

To increase the phase-space density and achieve quantum degeneracy, we perform forced evaporative cooling. The intensities of the dipole trap beams are lowered in an approximately exponential manner. The intensity of the horizontal dipole trap beam sets the trap depth and the intensity of the vertical dipole trap beam is used to tune the density of the sample. Evaporative cooling to quantum degeneracy takes between 550\,ms and 26\,s.

We now discuss some of the conditions required for successful evaporative cooling and the parameters with which we can influence these conditions. Evaporative cooling requires a high rate of elastic collisions to proceed faster than competing heating processes and loss of atoms. An important contribution to atom loss are three-body collisions. Both, elastic collisions and three-body collisions, depend on the scattering length $a$ and the density $n$. The different Sr isotopes and isotope mixtures span a large range of elastic scattering lengths [see Tab.~\ref{tab:SrScattering}]. These scattering lengths are not tunable in a way suitable for evaporative cooling, since magnetic Feshbach resonances do not exist and optical Feshbach resonances were shown to be accompanied by strong losses \cite{Chin2010fri,Ciurylo2005oto,Enomoto2008ofr,Blatt2011moo}. This situation leaves the density as the only practical parameter to increase the elastic collision rate and achieve a high ratio of elastic to three-body collisions. The elastic scattering rate scales as $a^2 n$, which favors large densities. The three-body collision rate has an upper limit proportional to $a^4 n^2$ \cite{Fedichev1996tbr,Bedaque2000tbr}. Therefore the ratio of elastic to three-body collisions is approximately $1/(a^2 n^2)$ and favors low densities, especially for the isotope with large scattering length, $^{86}$Sr. A limit to lowering the density is set by other loss processes and heating. Importantly, the optimum density depends also on the scattering properties of each isotope or isotopic mixture and is therefore tuned experimentally by adapting the power and the radius of the vertical dipole trap beam.

\subsection{Data acquisition and analysis}
\label{sec:ImprovedExperimentalDataAnalysis}

We characterize the gas by analysis of absorption images [see e.g. Fig.~\ref{fig:Fig4_LargeSr84BEC}(a)]. These images are taken in the ($\hat{\mathbf x}-\hat{\mathbf y}$)-direction [see Fig.~\ref{fig:Fig2_Setup}] after a free expansion time between 20\,ms and 30\,ms using the $^1S_0 - {^1P_1}$ transition. For high-density samples we detune the absorption imaging beam to reduce the optical density. The quadrupole magnetic field used for the MOT is switched off at the beginning of the free expansion. To analyze the images, we apply two-dimensional fits. For bosons, during the early stage of evaporation, the momentum distribution of the atom cloud is thermal and can in most cases be described by a single Gaussian. After the BEC phase-transition has occurred, we employ a bimodal fit to our data, consisting of a Thomas-Fermi distribution describing the BEC and a Gaussian for the thermal atoms \cite{Inguscio1999book}. The temperature of the sample is extracted from the Gaussian part of the fit. For fermions we use a Fermi-Dirac distribution, as in our previous work \cite{Tey2010ddb}. When working with a mixture of two isotopes, we only detect either one or the other of the isotopes after each experimental cycle. To avoid cross talk between the two isotopes from off-resonant absorption, we remove the isotope we do not want to image after 17\,ms of free expansion by a blast of resonant light on the very isotope-selective $^1S_0 - {^3P_1}$ intercombination line. To detect the internal state distribution of fermionic samples, we use the optical Stern-Gerlach technique \cite{Taie2010roa,Stellmer2011dam}. For situations in which the dipole trap potential can be well approximated by a harmonic potential, we calculate quantities of interest, as for example the peak density or the elastic collision rate, from trap oscillation frequencies, atom number, and temperature \cite{Inguscio1999book,Dalfovo1999tob,Pethick2002bec,Giorgini2008tou}.

\section{30-fold increase in $^{84}$Sr BEC atom number}
\label{sec:84BEC}

The first Sr isotope cooled to quantum degeneracy was $^{84}$Sr and BECs of $3\times10^5$ atoms have been obtained \cite{Stellmer2009bec,MartinezdeEscobar2009bec}. Here we increase this atom number by a factor 30, creating BECs exceeding $10^7$ atoms. To overcome the low natural abundance of $^{84}$Sr, we accumulate atoms in the metastable reservoir for 40\,s. This time is slightly longer than the lifetime of the gas in the reservoir, and further loading does not increase the atom number significantly. The atoms are returned into the ground state, cooled and compressed by the red MOT, and transferred into the dipole trap. For this experiment, we use only the horizontal dipole trap beam, which has an initial depth of $k_B \times 12(1)\,\mu$K and provides initial trapping frequencies of $f_x=45(5)\,$Hz, $f_y=6(1)\,$Hz, and $f_z=650(50)\,$Hz. After ramping the red MOT light off over 100\,ms, the gas is allowed to thermalize in the dipole trap for 250\,ms. At this point, about $4 \times 10^7$ atoms are in the dipole trap at a temperature of 1.5\,$\mu$K. The peak density of the gas is $7\times 10^{13}\,{\rm cm}^{-3}$, the average elastic collision rate $650\,{\rm s}^{-1}$, and the peak phase-space density 0.3. The power of the dipole trap is reduced exponentially from its initial value of 2.4\,W to 425\,mW within 10\,s.

\begin{figure}[htp]
\includegraphics[width=\columnwidth]{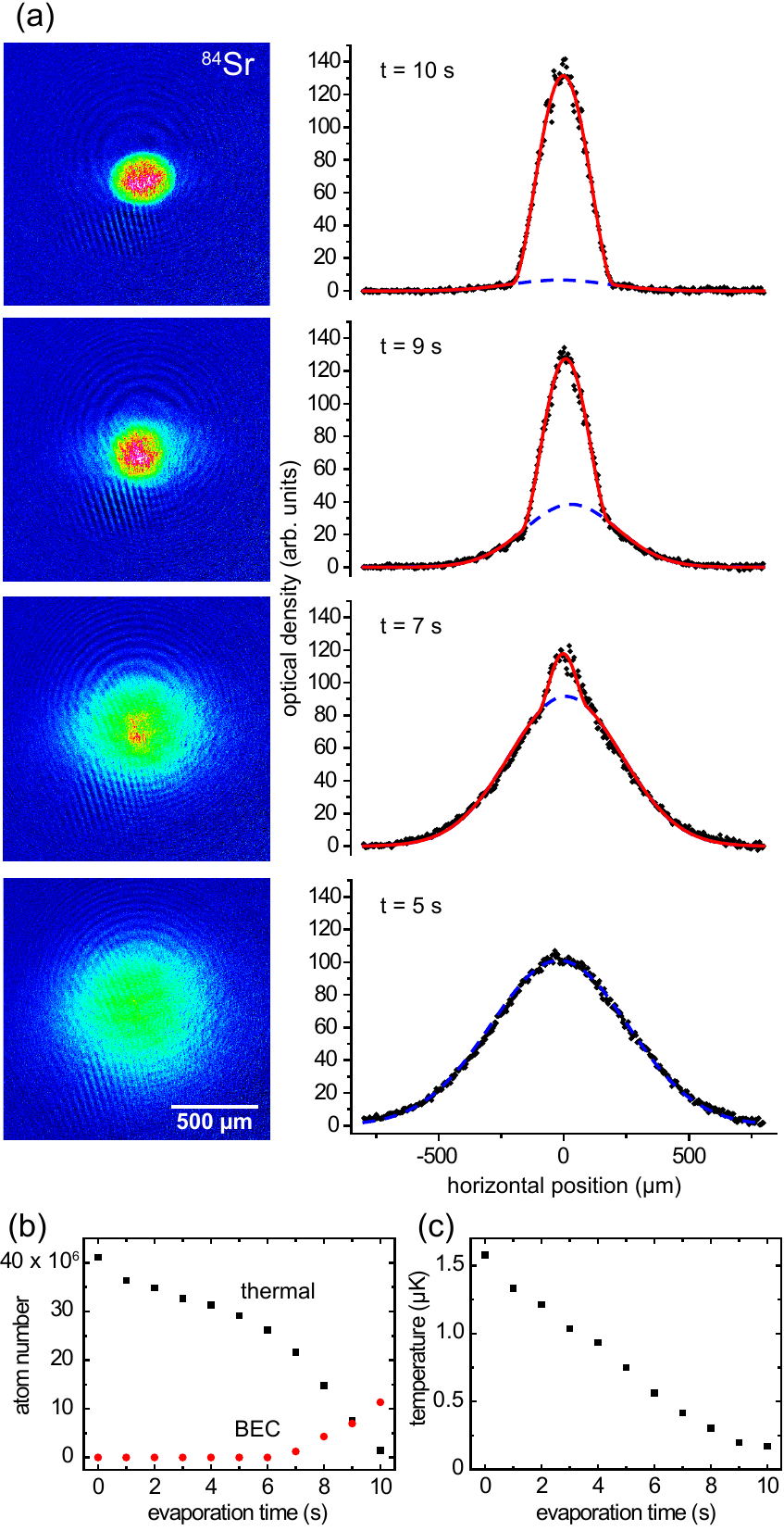}
\caption{\label{fig:Fig4_LargeSr84BEC} (Color online) Generation of $^{84}$Sr BECs exceeding $10^7$ atoms. Panel (a) shows a series of absorption images (left) and density profiles (right) for different times along the evaporative cooling ramp. The density profiles are obtained from the absorption images by integration along the vertical direction. The black circles are data points, which are fitted with a bimodal distribution (solid red line), capturing the thermal fraction (dashed blue line) and the BEC. Panel (b) shows the evolution of the atom number in the thermal (black squares) and condensate fraction (red circles) during evaporation. The temperature of the thermal fraction is shown in (c). }
\end{figure}

We investigate the behavior of the gas during evaporative cooling by careful analysis of absorption images taken after a free expansion time of 30\,ms [see Fig.~\ref{fig:Fig4_LargeSr84BEC}]. To reduce the optical density the imaging light is detuned by 48\,MHz from the $^1S_0 - {^1P_1}$ transition. Figure~\ref{fig:Fig4_LargeSr84BEC}(b) shows the evolution of thermal and BEC atom number and Fig.~\ref{fig:Fig4_LargeSr84BEC}(c) the evolution of temperature during evaporation. After 7\,s of evaporation a BEC is detected. At this time, $2.5\times10^7$ atoms remain in the trap at a temperature of about 400\,nK. The evaporation efficiency is high with four orders of magnitude gain in phase-space-density for a factor ten of atoms lost. After 10\,s of evaporation, we obtain an almost pure BEC of $1.1(1)\times 10^7$ atoms. The trap oscillation frequencies at this time are $f_x=20(3)\,$Hz, $f_y=2.5(5)\,$Hz, and $f_z=260(50)\,$Hz. The BEC has a peak density of $2.2\times 10^{14}\,{\rm cm}^{-3}$ and the shape of an elongated pancake with Thomas-Fermi radii of about $R_x=40\,\mu$m, $R_y=300\,\mu$m, and $R_z=3\,\mu$m. During the time of flight, the BEC expands by far the most in the vertical direction. Under the direction of imaging used, the shape of the resulting ellipsoid appears nearly circular [see Fig.~\ref{fig:Fig4_LargeSr84BEC}(a)]. The lifetime of the BEC is 15\,s, likely limited by three-body loss.

The atom number of this BEC is a factor 60 larger than in our previously reported work \cite{Stellmer2009bec}, and is to the best of our knowledge the largest BEC ever created by evaporative cooling in an optical dipole trap. The changes to our apparatus that led to this improvement are a higher oven flux, a reduction of the linewidth and an increase in the frequency accuracy of the red MOT cooling laser, an improved dipole trap design, a new laser source for the dipole trap with less intensity noise, and a careful optimization of the experimental sequence. It is remarkable that the Sr BEC with the largest atom number is created with the Sr isotope of lowest natural abundance.

An increase of the BEC atom number to nearly $10^8$ should be achievable by simple improvements. With more power available for the horizontal dipole trap, its horizontal waist could be increased while keeping the trap depth constant. The volume of the dipole trap would thus be increased and the trap oscillation frequency in the x-direction decreased. To obtain the same initial peak density, more atoms would have to be loaded into the dipole trap. This task could be achieved by improving the oven design to increase the atomic flux. An increased density of the final red MOT needs to be avoided to limit atom loss by light-assisted collisions, which could be done by using a larger detuning of the cooling laser, which increases the radial size of the red MOT [see Fig.~\ref{fig:Fig3_BoseFermiRedMOT}]. A limit of this strategy to increase the atom number occurs when the trap oscillation frequency in the x-direction becomes too low at the end of evaporation to permit the formation of a phase-coherent BEC during its lifetime. A decrease to the value currently used in the y-direction (2.5\,Hz) is certainly possible, which would already admit to increase the BEC atom number by a factor of six.

The large Sr BEC created here will increase the signal to noise in any study using $^{84}$Sr and is a promising coolant for other isotopes or atomic species.

\section{Generation of BECs with a cycle time of two seconds}
\label{sec:fastBEC}

\begin{figure}[htp]
\includegraphics[width=\columnwidth]{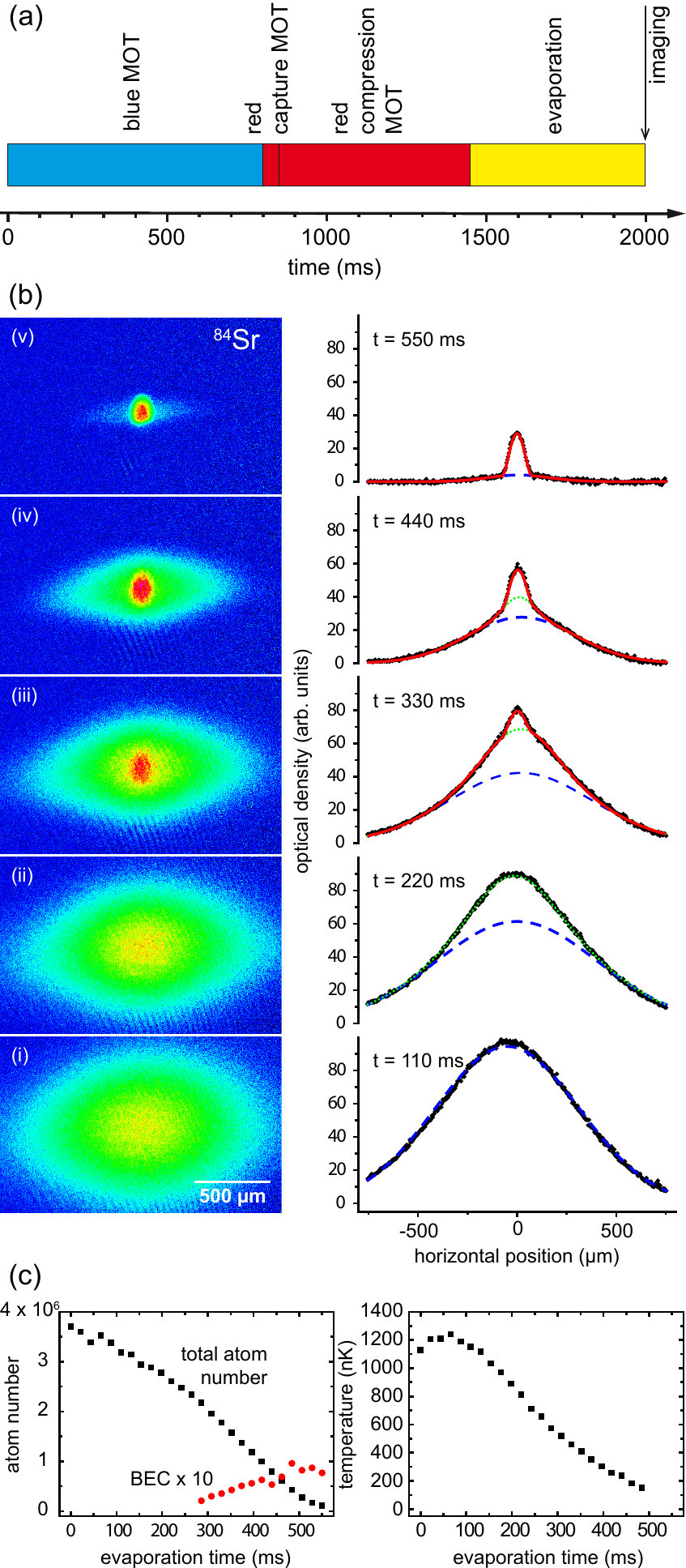}
\caption{\label{fig:Fig5_FastBEC} (Color online) Generation of $^{84}$Sr BECs with a short cycle time of the experimental sequence. Panel (a) shows a sketch of the experimental sequence. Absorption images taken after 22\,ms free expansion and integrated density profiles show the appearance of an an almost pure BEC of $10^5$ atoms at the end of evaporation (b). Dashed blue lines indicate Gaussians, dotted green lines double-Gaussians, and solid red lines bi- and trimodal distributions, see the text for details. Panel (c) shows the development of the total atom number and of the temperature corresponding to the narrower of the two Gaussian distributions during evaporation.}
\end{figure}

In the previous section, we have reported on experiments optimized for a large number of atoms in the BEC. We can also optimize our experimental sequence for a short cycle time. Nearly all experiments profit from the higher data rate made possible by a shorter cycle time. Precision measurement devices, such as atom interferometers, do require high repetition rates or a favorable ratio of probe time versus cycle time and might profit from the coherence of a BEC. Quantum gas experiments taking place in an environment of poor vacuum quality also require a short production time. Most experiments have cycle times of a few ten seconds. Experiments that have been optimized for speed while using an all-optical approach achieve cycle times of 3\,s for degenerate bosonic \cite{Kinoshita2005aob,Kraft2009bec} and 11\,s for fermionic quantum gases \cite{Duarte2011aop}. The same cycle time can be reached by using magnetic trapping near the surface of a microchip \cite{Farkas2010act,Horikoshi2006acb}.

Making use of the very high phase-space density achieved already in the red MOT, as well as the excellent scattering properties of $^{84}$Sr, we are able to reduce the cycle time to 2\,s [see Fig.~\ref{fig:Fig5_FastBEC}(a)]. At the beginning of the cycle, we operate the blue MOT for 800\,ms to load the metastable reservoir. A short flash of repump light returns the metastable atoms into the ground state, where they are trapped, compressed, and cooled to about 1.2\,$\mu$K by the red MOT. Close to $4\times10^6$ atoms are loaded into a dipole trap, which is formed by the horizontal sheet and a vertical beam of 25\,$\mu$m $1/e^2$-radius in the plane of the horizontal dipole trap. The atomic cloud is not only populating the cross of the dipole trap, but extends $\sim 1\,$mm along the horizontal dipole trap. Forced evaporation reduces the trap depth over 550\,ms with an exponential time constant of about 250\,ms.

Absorption images are taken after 22\,ms free expansion and analyzed as described in Sec.~\ref{sec:ImprovedExperimentalDataAnalysis} [see Fig.~\ref{fig:Fig5_FastBEC}(b)]. In this experiment we find that a single Gaussian is not sufficient to describe the thermal fraction. After about 100\,ms of evaporation, the distribution begins to deviate from a Gaussian without being close to quantum degeneracy [see Fig.~\ref{fig:Fig5_FastBEC}(b, case i)], indicating that part of the atoms reach a lower temperature than the rest of the sample. These are the atoms located in the crossing region of the horizontal and vertical beams, where the density is higher and evaporation is more efficient. After further evaporation, two samples with different temperatures are clearly visible, and we use a double Gaussian to fit our data (case ii). The phase transition occurs after about 270\,ms of evaporation, and we employ a trimodal fit to our data, describing the BEC, the thermal fraction of atoms in the crossing region, and the thermal atoms in the horizontal beam (cases iii and iv). After about 480\,ms of evaporation, the thermal fraction within the crossing region cannot be discerned, indicating an essentially pure BEC in this region. For this last stage of evaporation, we employ a bimodal fit to the data, capturing the BEC and the thermal atoms outside the crossing region (case v). Further evaporation does not increase the BEC atom number, but efficiently removes the thermal atoms. The evolution of atom number and temperature during evaporation is shown in Fig.~\ref{fig:Fig5_FastBEC}(c). About $10^5$ atoms reside in the BEC at the end of evaporation.

The read-out of the charged-coupled device (CCD) chip used for imaging can be performed during the consecutive experimental cycle and is therefore not included in the 2\,s period. The cycle time could be improved substantially if the reservoir loading time (800\,ms in this experiment) was reduced, e.g.~by increasing the oven flux. It seems that cycle times approaching 1\,s are within reach.

\section{Fivefold increase in $^{86}$Sr BEC atom number}
\label{sec:86BEC}

Some isotopes of alkaline-earth atoms feature large positive scattering lengths, such as $^{40}$Ca, $^{42}$Ca, $^{44}$Ca \cite{Dammalapati2011sco}, and $^{86}$Sr. While scattering between atoms provides thermalization during evaporation, there is a downside of a very large scattering length $a$: Inelastic three-body losses have an upper limit proportional to $a^4$ \cite{Fedichev1996tbr,Bedaque2000tbr}, and can reduce the evaporation efficiency drastically. Magnetic Feshbach resonances, a widely used means to tune the scattering length in ultracold samples, are absent in the alkaline-earth species, and a different strategy to reach degeneracy despite the large scattering length is needed.

In previous work \cite{Stellmer2010bec}, we have shown degeneracy of $^{86}$Sr, which has a scattering length of about 800\,$a_0$ \cite{Martinezdeescobar2008tpp}. In that experiment, the crucial innovation was to perform evaporation at a comparatively low density of $3 \times 10^{12}\,{\rm cm}^{-3}$ in a dipole trap of large volume. Two-body collisions, vital for thermalization, scale proportional with the density $n$, while detrimental three-body collisions scale as $n^2$. At small enough densities, evaporation can be efficient even for large scattering lengths. The trap was oblate with initial trap frequencies of $f_x=30\,$Hz, $f_y=3\,$Hz, and $f_z=260\,$Hz. After 4.8\,s of evaporation a nearly pure BEC containing 5000 atoms were created.

\begin{figure}[htp]
\includegraphics[width=\columnwidth]{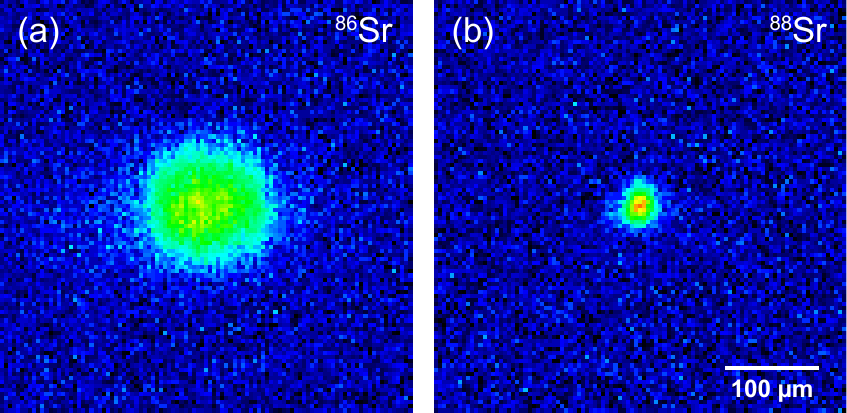}
\caption{\label{fig:Fig6_Sr86AndSr88BECs.pdf} (Color online) Essentially pure BECs of (a) $^{86}$Sr and (b) $^{88}$Sr. The free expansion time is 25\,ms for both images, and the geometric and color scales are identical. The BEC of $^{86}$Sr contains 25\,000 atoms and expands to much larger size compared to the almost non-interacting BEC of $^{88}$Sr, which contains 5000 atoms.}
\end{figure}

Improving this experiment, we are now increasing the vertical trap frequency by a factor two, which allows us to evaporate faster. Using a 500\,ms blue MOT stage, we load $9\times10^5$ atoms at a temperature of about 1\,$\mu$K into the dipole trap. The initial density is about $10^{12}\,{\rm cm}^{-3}$ and the average elastic collision rate 200\,s$^{-1}$. We perform evaporation much faster than in the previous $^{86}$Sr experiment, in just 800\,ms, which helps to avoid strong atom loss from three-body collisions. The onset of BEC is observed after 600\,ms of evaporation at a temperature of about 70\,nK with 350\,000\,atoms present. Further evaporation results in almost pure BECs of 25\,000\,atoms, which is an improvement of a factor five in atom number over our previous results [see Fig.~\ref{fig:Fig6_Sr86AndSr88BECs.pdf}(a)]. The cycle time of this experiment is again short, just 2.1\,s.

\section{Reliable generation of an attractive $^{88}$Sr BEC}
\label{sec:88BEC}

The isotope $^{88}$Sr is by far the most abundant one, and early research on cold gases of Sr focused on this isotope. The $s$-wave scattering length $a$ of $^{88}$Sr turned out to be very close to zero, which makes it an excellent choice for precision measurements \cite{Takamoto2005aol,Ferrari2006llb}. Optical clocks, for example, can suffer from density-dependent mean-field shifts, which scale $\propto a$. Similarly, force sensors based on Bloch oscillations \cite{Ferrari2006llb} suffer from dephasing by elastic collisions \cite{Gustavsson2008coi}. No other atomic species that has been laser-cooled so far combines the properties of high natural abundance, non-magnetic ground state, and very small interaction strength.

It is this small (and slightly negative) scattering length that hampered evaporative cooling of $^{88}$Sr towards degeneracy for a long time \cite{Ido2000odt,Poli2005cat}. Sympathetic cooling with the second-most abundant isotope, $^{86}$Sr, could not reach quantum degeneracy because of the large 3-body recombination rate of $^{86}$Sr \cite{Ferrari2006cos}.

Efficient evaporation to quantum degeneracy was first achieved by the Rice group \cite{Mickelson2010bec} using the fermionic isotope $^{87}$Sr as cooling agent. An ultracold gas of $^{88}$Sr is slightly attractive for $T\rightarrow0$, and indeed Mickelson and co-workers observed a limited number of atoms in the BEC, in good agreement with a simple mean-field model \cite{Ruprecht1995tds}.

Here, we confirm the work of the Rice group and highlight a few differences of our own experimental work. We start with an isotopic mixture in a crossed-beam dipole trap, containing $4.5\times10^6$ atoms of $^{88}$Sr at 560\,nK and $1.9\times10^6$ atoms of $^{87}$Sr at $1.0\,\mu$K. The sample thermalizes within the first second of forced evaporative cooling, and remains thermalized for the remainder of the 4-s ramp. The BEC of $^{88}$Sr appears after 2.1\,s, with $2.7\times10^6$ atoms present at a temperature of 320\,nK. Further evaporation does not increase the BEC atom number beyond a few thousand atoms, but the thermal fraction can be removed to obtain very pure BECs at the end of evaporation [see Fig.~\ref{fig:Fig6_Sr86AndSr88BECs.pdf}(b)].

A homogeneous BEC with negative scattering length is unstable for all atom numbers, but a trapped BEC can be stabilized through the zero-point kinetic energy of the trap up to a maximum atom number of $N_c=k\,a_{\mathrm{ho}}/a$ \cite{Dalfovo1999tob}. Here, $a_{\mathrm{ho}}=\sqrt{\hbar/(m\bar{\omega})}$ is the harmonic oscillator length, $\bar{\omega}=2\pi (f_x f_y f_z)^{1/3}$ and the prefactor $k\sim 0.55$ is only slightly dependent on the trapping geometry \cite{Gammal2001cno}. The maximum atom number thus depends only weakly on the trap frequencies, and one cannot expect substantial improvements from a change in trapping geometry. As the BEC atom number increases and eventually surpasses this limit during evaporation, part of the atoms are ejected out of the BEC \cite{Donley2001doc}, leaving the remnant behind to form a stable BEC again. This cycle continues throughout evaporation, until the remaining atom number is smaller than $N_c$ or the elastic collision rate is too low to reduce the temperature further and add atoms to the BEC. Consequently, one would expect a random atom number $N<N_c$ in the BEC at the end of evaporation. This behavior is close to the observation of Ref. \cite{Mickelson2010bec}: for a large number of experimental realizations, the BEC atom numbers were evenly distributed between $\sim 0.4\,N_c$ and $N_c$.

In view of the application of degenerate $^{88}$Sr in precision measurements, we will now show how the shot-to-shot variation in BEC atom number can be reduced. We deliberately start with a low initial number of $^{87}$Sr atoms to perform inefficient evaporation. After 80\% (or 3.5\,s) of evaporation, only 12\,000 $^{87}$Sr atoms remain, and the scattering rate of a thermal $^{88}$Sr atom with an $^{87}$Sr atom has dropped from initially 250\,s$^{-1}$ to below 4\,s$^{-1}$. At this point, we do not expect any appreciable cooling to take place. Instead, further lowering of the trap depth removes all thermal atoms from the trap, leaving behind a pure BEC of 6000 atoms. The atom number of this BEC is certainly below $N_c=7500$ for our conditions of $\bar{\omega}=2 \pi \times 50(5)$\,Hz and a scattering length of $a=-2\,a_0$ \cite{Stein2010tss}. The shot-to-shot variation in BEC atom number, however, is surprisingly small: We obtain a standard deviation of only 9\% in the atom number. This value is certainly larger than our typical $^{84}$Sr BEC atom number fluctuation of about 1\%, but it constitutes a substantial improvement towards precision measurements with $^{88}$Sr BECs. For a higher initial $^{87}$Sr atom number, we never observe a BEC exceeding $7000$ atoms. When identifying this number with $N_c$, we can deduce a scattering length of $-2.2(2)\,a_0$, which agrees well with the value derived from two-photon spectroscopy \cite{Stein2010tss}.

In conclusion, we have confirmed the $^{88}$Sr BEC experiments by the Rice group. We observe a much smaller variation in BEC atom number at the end of evaporation, which might be attributed to a slightly different choice of initial conditions.

\section{Double-degenerate Bose-Bose mixtures of Sr}
\label{sec:BoseBose}

\begin{figure}[htp]
\includegraphics[width=\columnwidth]{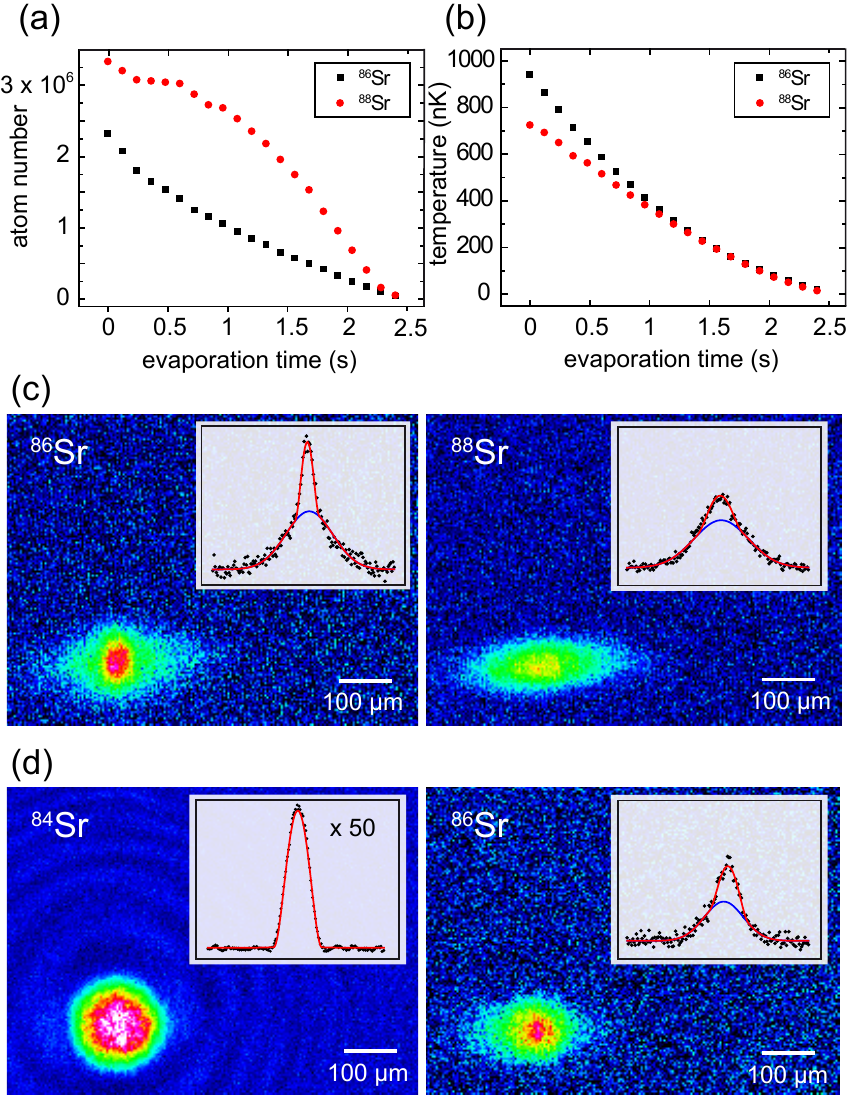}
\caption{\label{fig:Fig7_BoseBoseMixtures.pdf} (Color online) Creation of double-degenerate Bose-Bose mixtures. The development of atom number (a) and temperature (b) during evaporation are shown for the mixture of $^{86}$Sr (black squares) and $^{88}$Sr (red circles). Despite a significant temperature difference at the start of evaporation, the sample thermalizes within 1\,s. Below, we show absorption images of the mixtures $^{86}$Sr~+~$^{88}$Sr (c) and $^{84}$Sr~+~$^{86}$Sr (d), taken after 26\,ms of free expansion. The BEC of $^{84}$Sr is nearly pure and contains $2 \times 10^6$ atoms. The other three BECs are accompanied by a significant amount of thermal gas. The insets show the optical density integrated along the vertical direction, with bimodal fits to the data. Imaging of $^{84}$Sr is performed with a detuning of $30\,$MHz, the other isotopes are imaged on resonance. The color scale of the $^{84}$Sr image differs from the color scale used for the other three images, but the integrated optical densities in the insets are to scale.}
\end{figure}

Mixtures of two Bose-degenerate gases of different isotopes or elements allow the study of interesting phenomena, such as the miscibility and phase separation of two quantum fluids \cite{Hall1998doc,Riboli2002tot,Jezek2002ide} or the influence of one component on the other during the superfluid-to-Mott insulator transition \cite{Catani2008dbb}. They allow the application of isotope or species specific optical dipole potentials, which have been used to study entropy exchange between two bosonic gases \cite{Catani2009eei}, or the dynamics of impurities in a one-dimensional Bose gas \cite{Catani2012qdo}. Within the alkalis, a number of quantum degenerate Bose-Bose mixtures have been realized: $^{41}$K~+~$^{87}$Rb \cite{Modugno2002tas}, $^{85}$Rb~+~$^{87}$Rb \cite{Papp2008tmi}, and $^{87}$Rb~+~$^{133}$Cs \cite{Lercher2011poa}.

The many bosonic isotopes of alkaline-earth-metal-like elements in principle allow the creation of many different Bose-Bose mixtures. Unfortunately, for many of these mixtures, the interaction properties are unfavorable to create large and stable BECs. The absolute value of the two intra- and the interspecies scattering length must not be too large to avoid rapid decay and phase separation, but large enough for thermalization. The intraspecies scattering lengths should not be strongly negative to permit the formation of detectably large BECs \cite{Fukuhara2009aof}. The scattering length of alkaline-earth-metal-like atoms can only be tuned by optical Feshbach resonances, which introduce losses \cite{Ciurylo2005oto,Enomoto2008ofr,Blatt2011moo}. These limitations reduce the number of possible binary mixtures considerably. In particular, all combination of bosonic Ca isotopes seem unfavorable, since all intraspecies scattering lengths of the most abundant Ca isotopes are quite large \cite{Kraft2009bec,Dammalapati2011sco}. In Yb, two out of five bosonic isotopes have large negative scattering lengths \cite{Kitagawa2008tcp}, excluding many possible combinations of isotopes. One remaining combination, $^{170}$Yb~+~$^{174}$Yb, has a large and negative interspecies scattering length. One of the two remaining combinations, $^{168}$Yb~+~$^{174}$Yb, has been brought to double degeneracy very recently, with 9000 atoms in the BEC of each species \cite{Sugawa2011bec}. The interspecies scattering length between these two isotopes is $2.4 (3.4)\,a_0$ and provides only minuscule interaction between the two. The three bosonic isotopes of Sr give rise to three different two-isotope mixtures [see Tab.~\ref{tab:SrScattering}]. Of these the mixtures, $^{84}$Sr~+~$^{88}$Sr suffers from a large interspecies scattering length.

\begin{table}[b]
\caption{Scattering lengths $a$ between the Sr isotopes, given in units of $a_0$. The values are averages of the values given in \cite{Martinezdeescobar2008tpp} and \cite{Stein2010tss}, and the uncertainty is a few $a_0$ except for the two very large values ($a>500\,a_0$), where the uncertainty is much larger. All mixtures of isotopes, of which double-degenerate samples are presented in this work, are marked in bold. The natural abundance is given in the last column.}
	\label{tab:SrScattering}
	\centering
		\begin{tabular*}{\columnwidth}{@{\extracolsep{\fill}}cccccc}\hline \hline \noalign{\smallskip}
			          &$^{84}$Sr         & $^{86}$Sr         & $^{87}$Sr         & $^{88}$Sr     &  abundance (\%)    \\
\noalign{\smallskip}\hline\noalign{\smallskip}
			$^{84}$Sr & \textbf{123}     & \textbf{32}       & \textbf{-57}      & 1700          &         0.56      \\
			$^{86}$Sr & \textbf{32}      & \textbf{800}      & \textbf{162}      & \textbf{97}   &         9.86      \\
			$^{87}$Sr & \textbf{-57}     & \textbf{162}      & \textbf{96}       & \textbf{55}   &         7.00      \\
			$^{88}$Sr & 1700             & \textbf{97}       & \textbf{55}       & \textbf{-2}   &         82.58     \\ \hline \hline
		\end{tabular*}
\end{table}

Here we report on double-degenerate Bose-Bose mixtures of the combinations $^{84}$Sr~+~$^{86}$Sr and $^{86}$Sr~+~$^{88}$Sr, which have interspecies scattering lengths of $32\,a_0$ and $97\,a_0$, respectively. The experimental realization is straightforward: We consecutively load the two isotopes into the reservoir, repump them simultaneously on their respective $^3P_2-{^3D_2}$ transitions, and operate two red MOTs simultaneously. The mixture is loaded into the dipole trap and subsequently evaporated to form two BECs. Imaging is performed on the blue $^1S_0-{^1P_1}$ transition, and we image only one isotope per experimental run. The frequency shift between the isotopes is only about 4.5 linewidths. To avoid a contribution of the unwanted isotope to the absorption image, we remove the unwanted species by an 8-ms pulse of resonant light on the very isotope selective $^1S_0-{^3P_1}$ intercombination transition. To avoid a momentum distribution change of the imaged species by interspecies collisions, the pulse of light is applied after 17\,ms of free expansion, when the density of the sample has decreased significantly.

The experimental results are shown in Fig.~\ref{fig:Fig7_BoseBoseMixtures.pdf}. We will discuss the $^{86}$Sr~+~$^{88}$Sr combination first: $2.3\times 10^6$ ($3.3\times 10^6$) atoms of $^{86}$Sr ($^{88}$Sr) are loaded into the dipole trap, consisting of the horizontal beam and a weak vertical beam for additional axial confinement. The initial temperatures of the two species are quite different: 950\,nK for $^{86}$Sr and 720\,nK for $^{88}$Sr, which reflects the different intraspecies scattering behavior. The interspecies scattering length is around $100\,a_0$, and the two species clearly thermalize to reach equilibrium after 1\,s of evaporation. As the trap depth is lowered further, we observe the onset of BEC in $^{86}$Sr ($^{88}$Sr) after 2.0\,s (2.3\,s). At the end of our evaporation ramp, which lasts 2.4\,s, we obtain 10\,000 (3000) atoms of $^{86}$Sr ($^{88}$Sr) in the condensate fraction [see Fig.~\ref{fig:Fig7_BoseBoseMixtures.pdf}(a)]. Further evaporation does not increase the BEC atom numbers.

In a second experiment, we investigate the $^{84}$Sr~+~$^{86}$Sr mixture with an interspecies scattering length of $32\,a_0$. Starting out with $10\times 10^6$ ($1.5\times 10^6$) atoms of $^{84}$Sr ($^{86}$Sr) in the dipole trap, we perform forced evaporation over 2\,s, and the two species remain in perfect thermal equilibrium throughout this time. The phase transition of $^{84}$Sr is observed already after 1.3\,s, with about $2.5 \times 10^6$ atoms present at a temperature of 200\,nK. After 1.9\,s, the BEC is essentially pure and contains up to $2\times 10^6$ atoms. The atom number of $^{86}$Sr is kept considerably lower to avoid three-body loss. The phase transition occurs later: after 1.7\,s, with $4 \times 10^5$ atoms at a temperature of 130\,nK. Till the end of evaporation, the BEC fraction grows to 8000 atoms but remains accompanied by a large thermal fraction [see Fig.~\ref{fig:Fig7_BoseBoseMixtures.pdf}(d)].

In conclusion, we have presented two binary Bose-Bose mixtures of alkaline-earth atoms with appreciable interaction between the two species. These mixtures enjoy the property that isotope-selective optical traps can be operated close to one of the intercombination lines. This might allow for an individual addressing of the isotopes by a dipole trap operated close to these transitions \cite{Yi2008sda}, reminiscent of the case of Rb in its hyperfine states $|F=1\rangle$ and $|F=2\rangle$ \cite{Mandel2003cto} or nuclear substates in Yb and Sr \cite{Taie2010roa,Stellmer2011dam}.

\section{Deeply-degenerate Fermi Gases of $^{87}$Sr}
\label{sec:87DFG}

A wealth of recent proposals suggest to employ the fermionic isotopes of alkaline-earth-metal-like atoms as a platform for the simulation of SU($N$) magnetism \cite{Wu2003ess,Wu2006hsa,Cazalilla2009ugo,Hermele2009mio,Gorshkov2010tos,Xu2010lim,FossFeig2010ptk,FossFeig2010hfi,Hung2011qmo,Manmana2011smi,Hazzard2012htp,Bonnes2012alo}, for the generation of non-Abelian artificial gauge fields \cite{Dalibard2011agp,Gerbier2010gff}, to simulate lattice gauge theories \cite{Banerjee2012aqs}, or for quantum computation schemes \cite{Hayes2007qlv,Daley2008qcw,Gorshkov2009aem,Daley2011sdl}.

Elements with a large nuclear spin are especially well suited for some of these proposals. They allow to encode several qubits in one atom \cite{Gorshkov2009aem}, and could lead to exotic quantum phases, as chiral spin liquids, in the context of SU($N$) magnetism \cite{Hermele2009mio}. Furthermore, it has been shown that the temperature of a lattice gas is lower for a mixture containing a large number of nuclear spin states after loading the lattice from a bulk sample \cite{Hazzard2012htp,Taie2012asm,Bonnes2012alo}. The largest nuclear spin of any alkaline-earth-metal-like atom is 9/2, and it occurs in the nuclei of $^{87}$Sr and of two radioactive nobelium isotopes. This fact makes $^{87}$Sr with its ten spin states an exceptional candidate for the studies mentioned.

\begin{figure}[htp]
\includegraphics[width=\columnwidth]{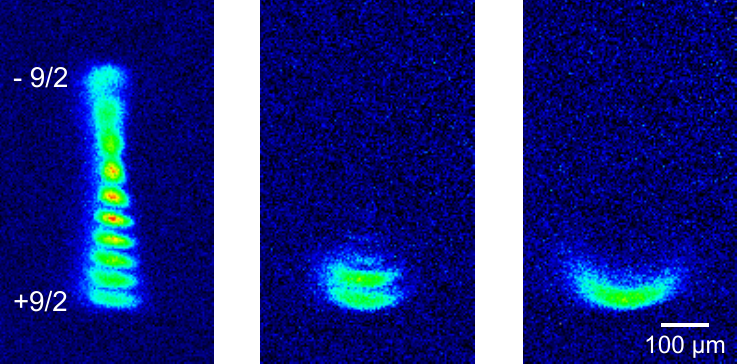}
\caption{\label{fig:Fig8_OSG.pdf} (Color online) Detection of spin state distribution using the optical Stern-Gerlach technique \cite{Taie2010roa,Stellmer2011dam}. Samples of $^{87}$Sr in a ten-state mixture or optically pumped into two or one spin states are shown.}
\end{figure}

\begin{figure}[htp]
\includegraphics[width=\columnwidth]{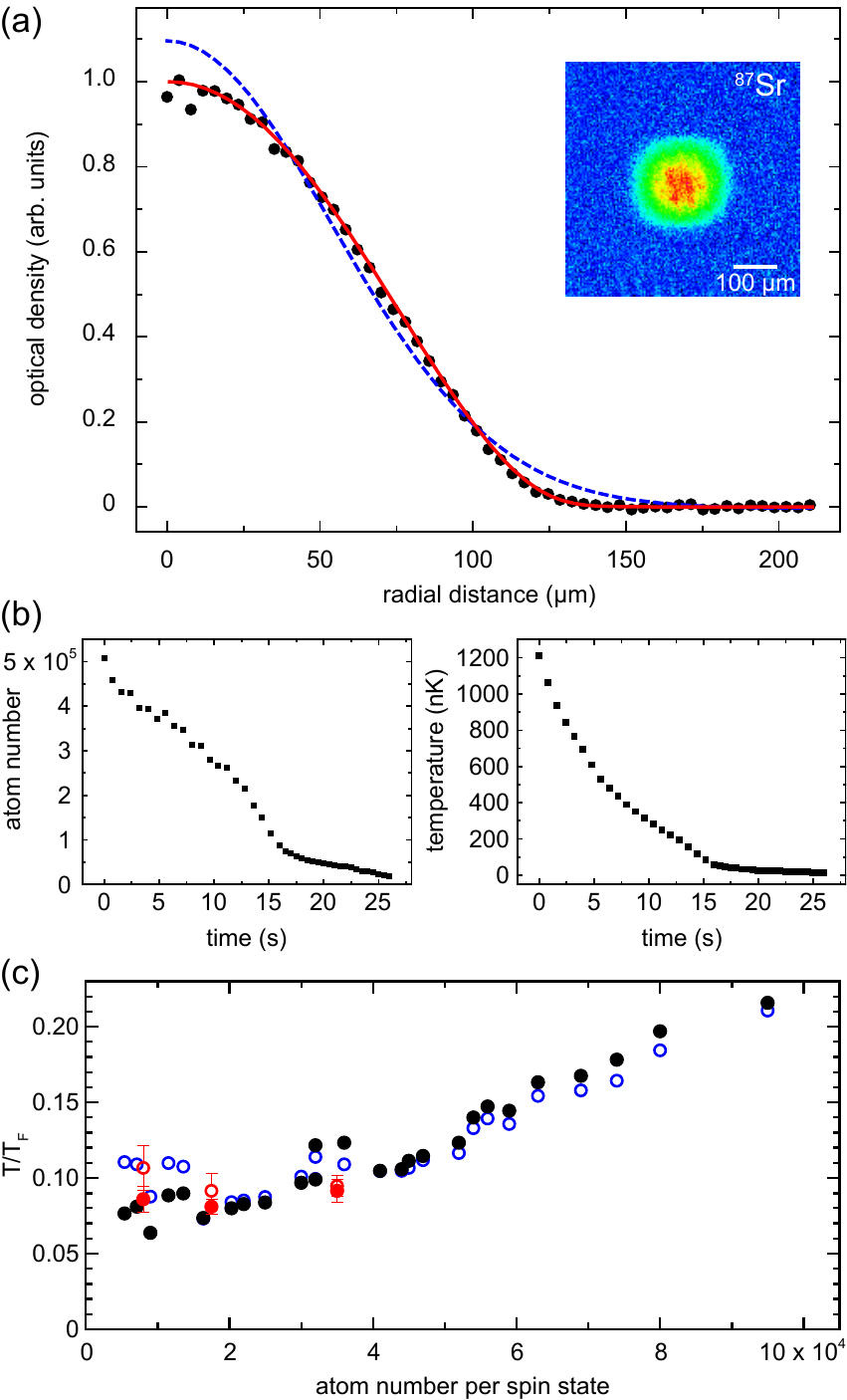}
\caption{\label{fig:Fig9_Sr87DFG.pdf} (Color online) Deeply degenerate Fermi gases of $^{87}$Sr in a balanced mixture of ten nuclear spin states. Panel (a) shows the azimuthally averaged density distribution of a degenerate Fermi gas at $T/T_F=0.08(1)$ after 25.4\,ms of free expansion (black circles). The measurement is well described by a Fermi-Dirac distribution (red line) but not by a Gaussian (dashed blue line). The corresponding absorption image is shown in the inset. Panel (b) shows the evolution of the atom number per spin state and of the temperature during evaporation, derived from two-dimensional fits of Fermi-Dirac distributions to time-of-flight absorption images. Panel (c) shows the value of $T/T_F$ in dependence of atom number per spin state, derived from the fugacity (open circles) or calculated from the temperature, the atom number, and the trap oscillation frequencies (filled circles). Some data points have been taken multiple times (red circles with error bars) to determine the statistical uncertainty. }
\end{figure}

The study of SU($N$) magnetism in a lattice requires the temperature of the sample to be below the super-exchange scale, $t^2/U$, where $t$ is the tunnel matrix element and $U$ the on-site interaction energy \cite{Bonnes2012alo}. A high degree of degeneracy in the bulk would constitute a good starting point for subsequent loading of the lattice. A value of $T/T_F=0.26(5)$ in a ten-state mixture \cite{DeSalvo2010dfg} and of $T/T_F=0.30(5)$ in a single-spin state \cite{Tey2010ddb} of $^{87}$Sr were presented already, both with about 20\,000 atoms in each Fermi sea. Neither of these two experiments had the means to set the number of spin states at will. Here, we present degenerate Fermi gases of Sr at considerably deeper degeneracy with an arbitrary number of spin states between 1 and 10.

About $5\times10^6$\,atoms of $^{87}$Sr are loaded into the optical dipole trap, where we measure a temperature of $1.2\,\mu$K. At this time, atoms are in a roughly even mixture of all spins. To prepare the desired spin mixture, we perform optical pumping on the $^1S_0-{^3P_1}, |F=9/2\rangle \rightarrow |F'=9/2\rangle$ transition at a small guiding field of 3\,G, which splits adjacent $m_{F'}$-states by 260\,kHz, corresponding to 35 linewidths. As we have shown before \cite{Stellmer2011dam}, we can prepare any combination and relative population of the ten spin states. The optical pumping is optimized using the optical Stern-Gerlach technique [see Fig.~\ref{fig:Fig8_OSG.pdf}] \cite{Taie2010roa,Stellmer2011dam}, and quantified using state-selective absorption imaging on the intercombination line \cite{Stellmer2011dam}. We can reduce the population of undesired spin states to below 0.1\%, where this value is limited by our detection threshold of 3000 atoms. After the spin preparation, which does not heat the sample, we perform evaporative cooling. Evaporation proceeds in two stages. The first stage lasts 16\,s, during which the power of the horizontal dipole trap beam is reduced by a factor of five. At this point, the gas enters the degenerate regime with a typical temperature of around $0.3\,T_F$ and evaporative cooling becomes less efficient because of Pauli blocking \cite{Demarco2001pbo}. We compensate for this effect by a reduced ramp speed during the second evaporation stage. The power of the horizontal beam is only reduced by a factor 1.5 during 10\,s. Slightly different final trap depths are used for different numbers of populated spin states. Trap frequencies at the end of evaporation are $f_x\sim30\,$Hz, $f_y\sim30\,$Hz, and $f_z\sim200\,$Hz. The case of a single populated spin state is unique since identical fermions do not collide at low temperature because of symmetry requirements, impeding thermalization of the sample. In this case, we add $^{84}$Sr atoms to the system to facilitate sympathetic cooling.

In all of these experiments, we determine the atom number, temperature $T$, and Fermi temperature $T_F$ by fitting two-dimensional Fermi-Dirac distributions to absorption pictures as in our previous work \cite{Tey2010ddb}. Two methods are used to determine $T/T_F$. Either $T$ is determined by the fit and $T_F$ is calculated from the atom number $N_{\rm at}$ and average trap frequency as $T_F=\hbar \bar\omega (6N_{\rm at})^{1/3}/k_B$. Or $T/T_F$ is calculated directly from the fugacity, which is a fit parameter. The momentum distribution of a gas at small $T/T_F$ strongly deviates from a Gaussian shape, which we show in azimuthally integrated profiles [see Fig.~\ref{fig:Fig9_Sr87DFG.pdf}(a)].

We will limit the presentation of our data to the cases of $N=10$, $N=2$, and $N=1$. For the ten-state mixture, the fraction of Pauli-forbidden collisions is small, and evaporative cooling performs well to yield a stack of ten spatially overlapping Fermi seas. Figures~\ref{fig:Fig9_Sr87DFG.pdf}(b) show the development of atom number per spin state and temperature along the evaporation ramp. Figure~\ref{fig:Fig9_Sr87DFG.pdf}(c) shows $T/T_F$ in dependence of atom number. With about 30\,000 atoms per spin state, we obtain $T/T_F=0.10(1)$ at $T_F=160\,$nK. The errors given here are statistical errors of multiple experimental realizations, and we estimate systematic errors to be of similar magnitude.

For the two-state mixture, we pump all atoms into a balanced population of $m_F=+9/2$ and $m_F=+7/2$ states. The total atom number, initial temperature, and evaporation trajectory are identical to the previous case of a ten-state mixture, but there is a crucial difference: Only half of all collisions possible for distinguishable particles are Pauli-allowed in the binary mixture, leading to a decreased thermalization rate. The reduced evaporation efficiency immediately manifests itself in the degree of degeneracy reached: Despite the higher atom number per spin state, we can reach only $T/T_F=0.20(1)$ with $1.0\times 10^5$\, atom per spin state and $T_F=230\,$nK. The lowest value reached is $T/T_F=0.17(1)$ with 60\,000 atoms remaining. Further evaporation does not reduce $T/T_F$.

In the case of a single-spin sample of $2.5\times 10^6$ fermionic atoms, we add $6.5\times 10^6$ atoms of $^{84}$Sr to the dipole trap. Evaporation is performed in one single exponential ramp over 8\,s and results in a pure BEC of $7\times 10^5$\,atoms. Towards the end of evaporation, the fermionic cloud contains $10^5$\,atoms. The sample is not well thermalized in the axial direction of the trap, which we take into account by fitting the vertical and horizontal directions independently to obtain $T/T_F=0.12$ and $T/T_F=0.23$, respectively.

Comparing the ten-state spin mixture to previously published results \cite{DeSalvo2010dfg,Tey2010ddb}, we reach a $\sim 2.5$ times lower value of $T/T_F$ at a comparable atom number per spin state. A deeply degenerate gas is a crucial requirement for the reliable operation of a quantum computer based on alkaline-earth-metal-like atoms or the exploration of SU($N$) magnetism.

\section{Quantum degenerate Bose-Fermi mixtures of Sr}
\label{sec:BoseFermi}

\begin{figure}[htp]
\includegraphics[width=\columnwidth]{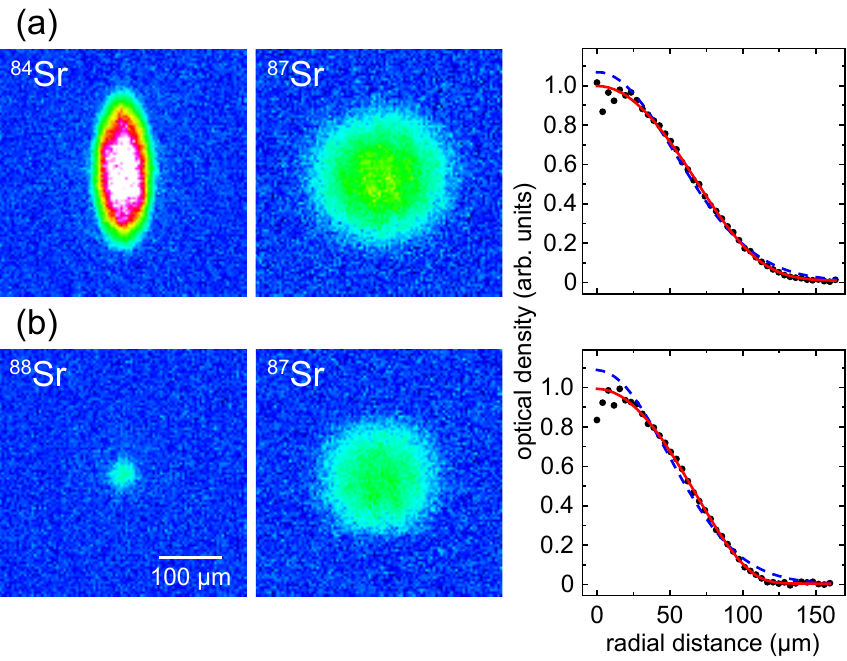}
\caption{\label{fig:Fig10_BoseFermiMixtures.pdf} (Color online) Double-degenerate Bose-Fermi mixtures. The two combinations shown here are $^{84}$Sr~+~$^{87}$Sr (a) and $^{88}$Sr~+~$^{87}$Sr (b), where all ten nuclear spin states of $^{87}$Sr are populated. The absorption images of the bosonic (left column) and fermionic isotopes (middle column) are taken after 25.4\,ms of free expansion. The color and length scales are identical for all four images. Azimuthally integrated optical density distributions of the fermionic cloud are shown in the right column, together with Fermi-Dirac (solid red line) and Gaussian fits (dashed blue line).}
\end{figure}

Double-degenerate Bose-Fermi mixtures of ultracold gases have been realized with alkali-metal mixtures \cite{Truscott2001oof,Schreck2001qbe,Roati2002fbq,Hadzibabic2002tsm,Wu2011sii,Park2012qdb}, Li~+~Yb mixtures \cite{Hara2011qdm,Hansen2011qdm}, and isotopic mixtures of Yb \cite{Fukuhara2009aof}, Sr \cite{Tey2010ddb}, and metastable He \cite{McNamara2006dbf}. Here, we introduce three new quantum degenerate mixtures. We combine each of the three bosonic Sr isotopes with a ten-spin-state mixture of $^{87}$Sr. The distinguishing property of the three mixtures are the different inter- and intraspecies scattering lengths. The interspecies scattering length of $^{84}$Sr, $^{86}$Sr, and $^{88}$Sr with $^{87}$Sr are $-57\,a_0$, $162\,a_0$, and $55\,a_0$, respectively, and all allow efficient interspecies thermalization.

In a first experiment, we prepare the mixture of $^{84}$Sr~+~$^{87}$Sr. In contrast to the experiments presented in Ref.~\cite{Tey2010ddb} and Sec.~\ref{sec:87DFG}, which used spin-polarized fermions, we are now working with the entire ten-state mixture of $^{87}$Sr. About $6\times10^6$ fermions at $1.15\,\mu$K and $2.1\times10^6$ bosons at $1.1\,\mu$K are loaded into the dipole trap. The first of two exponential evaporation ramps takes 12\,s, during which the onset of BEC in $^{84}$Sr is observed after 8\,s. The BEC is essentially pure after 12\,s, containing about $3\times10^5\,$atoms. The fermions remain in perfect thermal equilibrium with the bosons. At the end of this first ramp, we obtain $7\times10^4$ fermions per spin state at a temperature of about 100\,nK, but still well outside the degenerate regime. For further cooling, we add a very slow second evaporation ramp of 7.5\,s duration. Such a slow ramp is required because Pauli blocking decreases the scattering rate between the fermions, and superfluidity of the BEC decreases the scattering rate between fermions and bosons. The final trap frequencies are $f_x=f_y=30\,$Hz, and $f_z=220\,$Hz. The degree of Fermi degeneracy increases substantially to $T/T_F=0.15(1)$ with 15\,000 atoms in each spin component at the end of evaporation. The BEC atom number decreases slightly to $2\times10^5\,$atoms [see Fig.~\ref{fig:Fig10_BoseFermiMixtures.pdf}(a)]. This experiment reaches an eleven-fold degeneracy of distinguishable particles, possibly the largest number of overlapping degenerate gases ever reported. It seems that the presence of the bosons reduces the evaporation efficiency of the fermions, as deeper degeneracies were achieved in absence of the bosons; see Sec.~\ref{sec:87DFG}.

In a second experiment, we use the bosonic isotope $^{86}$Sr. The bosonic intraspecies and the interspecies scattering lengths are much larger than in the previous case, which we account for by decreasing the density of the sample. We load less atoms ($1.6\times10^6$\,atoms of $^{86}$Sr at $1.1\,\mu$K and $3.7\times10^6$\,atoms of $^{87}$Sr at $1.0\,\mu$K) and we keep the average trap frequency low by reducing the horizontal confinement. We maintain the concept of two sequential evaporation ramps of different time constants and evaporate slightly deeper than is the previous case, but reduce the total evaporation time to 3.4\,s. Final trap frequencies are $f_x=20\,$Hz, $f_y=10\,$Hz, and $f_z=200\,$Hz. This isotopic combination performs worse than the previous one, yielding a bosonic sample with only 15\% condensate fraction. The horizontal trap frequencies, required to be small to keep three-body loss of $^{86}$Sr low, does not ensure thermalization of the fermionic sample in this direction. We obtain $T/T_F=0.15(5)$ for the vertical direction. The BEC contains 5000 atoms, and each Fermi sea contains 10\,000 atoms.

As a last experiment, we use the $^{88}$Sr isotope as the boson. Starting out with $1.2\times10^6$ bosons and $6.5\times10^6$ fermions both at $1.2\,\mu$K, we reduce the trap depth in two ramps of 12\,s and 8\,s. The atom number of the $^{88}$Sr BEC is limited by the negative scattering length, and evaporation to a low trap depth is required to remove the thermal fraction. We finally obtain a pure BEC of 4000\,atoms immersed in ten Fermi seas, each comprising 10\,000\,atoms at $T/T_F=0.11(1)$ [see Fig.~\ref{fig:Fig10_BoseFermiMixtures.pdf}(b)].

\begin{table*}[htbp!]
	\centering
\caption{Overview of the Sr quantum gases and quantum gas mixtures discussed in this article. The isotopes used in a certain experiment are designated as ``isotope 1'' and ``isotope 2''. If the fermionic $^{87}$Sr isotope is used, the number of populated nuclear spin states is given. The initial conditions of the dipole trap are described by the $1/e^2$-radius of the vertical dipole trap beam in the plane of the horizontal dipole trap $w_{\rm vert}$ and the power of the horizontal and vertical dipole trap beams $P_{\rm (horiz/vert), start}$. The waists of the horizontal dipole trap beam are the same for all experiments, 250(50)\,$\mu$m in the x-direction and 18(2)\,$\mu$m in the z-direction. The dipole trap depth is $U=k_B \times P_{\rm horiz, start} \times 5.0(5)\,\mu$K/W. The reservoir loading time of isotope $i$ is given by $t_{{\rm load}, i}$. After transfer of atoms from the red MOT into the dipole trap, the sample contains $N_i$ atoms at temperature $T_i$ of isotope $i$. Evaporative cooling is performed using one or two nearly exponential evaporative cooling ramps with duration $t_{\rm evap, (1/2)}$. Two ramps are used to create degenerate Fermi gases (DFGs). The second ramp has a much longer time constant than the first ramp, which accounts for the reduction of evaporation efficiency by Pauli blocking for samples in the quantum degenerate regime. The powers of the dipole trap beams at the end of evaporation are given by $P_{\rm (horiz/vert), end}$. At this point the trap is well approximated by a harmonic potential with trap oscillation frequencies $f_{x,y,z}$. The last lines of the table characterize the resulting quantum gases, their atom number $N$, the phase transition temperature $T_c$ for bosonic isotopes, $T/T_F$ and Fermi temperature $T_F$ for the fermionic isotope, and the peak densities of the gas $n_0$. All errors are estimated to be below 20\,\% unless stated otherwise.}
	\label{tab:Summary}
		\begin{tabular*}{\textwidth}{@{\extracolsep{\fill}}llcccccccccccc}
\hline\hline \noalign{\smallskip}
& & \multicolumn{4}{c}{BEC} & \multicolumn{2}{c}{BEC+BEC} & \multicolumn{3}{c}{DFG} & \multicolumn{3}{c}{DFG+BEC}  \\
\cmidrule[0.3pt](r){3-6}\cmidrule[0.3pt](lr){7-8}\cmidrule[0.3pt](lr){9-11}\cmidrule[0.3pt](l){12-14}\\
isotope 1 & & $^{84}$Sr   & $^{84}$Sr   &  $^{86}$Sr   & $^{88}$Sr   & $^{86}$Sr   & $^{84}$Sr   & $^{87}$Sr   & $^{87}$Sr   & $^{87}$Sr   & $^{87}$Sr   & $^{87}$Sr   & $^{87}$Sr  \\
isotope 2 & & ---         & ---         & ---          & $^{87}$Sr   & $^{88}$Sr   & $^{86}$Sr   & ---         & ---         & $^{84}$Sr   & $^{84}$Sr   & $^{86}$Sr   & $^{88}$Sr \\ \multicolumn{2}{l}{\# fermionic spin states}& ---       & ---         &  ---         &  10         &  ---        &  ---        &  10         &  2          & 1           & 10          & 10          & 10       \\
\noalign{\smallskip}\hline\noalign{\smallskip}
$w_{\rm vert}$ &($\mu$m) &  ---  & 25         & 230         & 60         &  60         & 230          & 60          & 60        & 60        & 60        & 60        &   60    \\
$P_{\rm horiz, start}$& (mW) & 2400 & 2000 & 1000 & 2000 & 1700 & 1200 & 2000 & 2000 & 2000 & 2000 & 2000 & 2000 \\
$P_{\rm vert, start}$&(mW) &  0 & 185 & 3 & 180 & 3 & 20 & 220 & 180 & 180 & 120 & 0 & 120 \\
\noalign{\smallskip}\hline\noalign{\smallskip}
$t_{\rm load, 1}$& (s) & 40     & 0.8     & 0.5          & 0.075   & 0.2   & 5   & 20         & 10         & 10   & 10   & 2.5   & 10   \\
$t_{\rm load, 2}$& (s) & ---     & ---     & ---          & 10   & 0.05   & 0.3   & ---         & ---         & 10   & 1   & 0.1   & 0.05   \\
$N_1$ &($10^6$) &$40$&$4$&$0.9$&$4.5$&$2.3$&$10$&$5$&$5$&$5$&$6$&$3.7$&$6.5$ \\
$N_2$ &($10^6$)&---&---&---&$1.9$&$3.3$&$1.5$&---&---&$6.5$&$2.1$&$1.6$&$1.2$ \\
$T_1$ &($\mu$K) &1.5&1.2&$\sim 1$&0.56&0.95&$\sim 1$&1.2&1.2&1.2&1.15&1.0&1.2 \\
$T_2$ &($\mu$K) & ---&---&---&1&0.72&$\sim 1$&---&---&1.2&1.1&1.1&1.2 \\
\noalign{\smallskip}\hline\noalign{\smallskip}
$t_{\rm evap, 1}$ &(s) & 10 & 0.55 & 0.8 & 3 & 2.4 & 2 & 16 & 16 & 8 & 12 & 2.4 & 12\\
$t_{\rm evap, 2}$ &(s) & --- & --- & --- & --- & --- & --- & 10 & 10 & --- & 7.5 & 0.8 & 8\\
$P_{\rm horiz, end}$&(mW) & 425 &265 &215 &275 &295 &185 &345 &295 &265 &345 &325 &265  \\
$P_{\rm vert, end}$&(mW) & 0 & 15 & 40 & 30 & 3 & 20 & 30 & 30 & 20 & 30 & 5 & 30 \\
$f_{x}$& (Hz)      & 20  & 120  & 20  & 30  & 20  & 20  & 30  & 30  & 25  & 30  & 20  & 30 \\
$f_{y}$ &(Hz)      & 2.5  & 110  & 2.5  & 30  & 10  & 3.5  & 30  & 30  & 25  & 30  & 10 & 30 \\
$f_{z}$ &(Hz)      & 260  & 170  & 110  & 160  & 180  & 130  & 220  & 180  & 150 & 220  & 200  & 150 \\
\noalign{\smallskip}\hline\noalign{\smallskip}
$N_{\rm BEC\,1}$ &($10^3$)& 11000 & 100 & 25 & 6 & 10 & 2000 & --- & --- & --- & --- &--- & --- \\
$N_{\rm BEC\,2}$ &($10^3$) & --- & --- & --- & --- & 3 & 8 & --- & ---  & 700 & 200 & 5 & 4 \\
$N_{\rm DFG}$ per state & ($10^3$)  & --- & --- & --- & --- & --- & --- & 30 & 100 & 100 & 15 & 10 & 10  \\
$T_{c,1}$& (nK) & $\sim 400$ & 580 & 70 & 320 & 80 & 200 & --- & --- & --- & --- & --- & --- \\
$T_{c,2}$ &(nK) &  --- &  --- &  --- &  --- & 30 & 130 &  --- &  --- & 560 & 220 & 60 & 170 \\
$T_F$& (nK) &  ---        & ---         & ---         & ---         &  ---        &  ---     &  160        &  220        & 180         &   130       &  70        &  100  \\
$T/T_F$ & & --- & --- & --- & --- & --- & --- & 0.10(1) & 0.20(1) & 0.12/0.23 & 0.15(1) & 0.15(5) & 0.11(1) \\
$n_{\rm 0, BEC\,1}$ &($10^{12}\,$cm$^{-3}$) &   220       &  260    & 4         &  ---     &  6        & 100         & ---        &  ---  & ---      &   ---  &  ---   &    --- \\
$n_{\rm 0, BEC\,2}$ &($10^{12}\,$cm$^{-3}$) &   ---       &  ---    & ---         &  ---        &  ---        & 3     & ---     &  ---  & 170    &  140  &  6   &    --- \\
$n_{\rm 0, DFG}$ per state &($10^{12}\,$cm$^{-3}$) &   ---       & ---         &  ---        & ---    &  ---  & ---     &  7        &  12  & 9      &   5  &  2   &  3   \\
\hline \hline
		\end{tabular*}
\end{table*}

\section{Conclusions}
\label{sec:conclusion}

We have presented a collection of experiments related to the creation of BECs and degenerate Fermi gases of Sr. We have shown how electronic and collisional properties add to create a strong framework delivering a diversity of quantum degenerate samples [see Tab.~\ref{tab:Summary}]. Key parameters of Sr quantum gases, such as the BEC atom number and the degeneracy of fermionic samples, have been improved significantly over previous experiments. In particular, we have presented the largest BECs ever created by evaporative cooling in an optical trap, we were able to reduce the experimental cycle time for BEC preparation to 2\,s, and we have prepared degenerate Fermi gases at $T/T_F=0.10(1)$.

Four key properties of Sr make these results possible. First, the 461-nm broad-linewidth transition enables efficient laser cooling. Second, atoms in the metastable $^3P_2$ state can be accumulated in a magnetic trap. This property allows us to overcome the low natural abundance of $^{84}$Sr. Third, the 689-nm intercombination line, which has a linewidth of only 7.4\,kHz, enables laser cooling to the exceptionally high phase-space density of 0.1. And finally, the collision properties of $^{84}$Sr and $^{87}$Sr are favorable for efficient evaporative cooling. Since our first results on quantum degenerate Sr, we have implemented a range of technical improvements to our apparatus, which allow us to make best use of these properties. Besides a reduced linewidth of the intercombination line cooling laser source and an overall optimization of the experimental sequence, the most important improvement is the design of our dipole trap. Its flexibility allows us to quickly choose optimum conditions for each Sr isotope or isotope mixture. Some of these techniques might also be applicable to other atoms. The metastable state in Yb could be used to accumulate atoms, but would require a dedicated laser pumping the atoms from the MOT cycle into the metastable state.

The work presented here did pave the ground for the creation of bosonic and fermionic Sr lattice gases \cite{StellmerPhD} and the creation of Sr$_2$ molecules \cite{Stellmer2012cou}. We plan to use the large $^{84}$Sr BEC as thermal bath for the creation of quantum degenerate Rb/Sr mixtures. The exploration of SU($N$) magnetism with Sr will greatly benefit from the highly degenerate fermionic quantum gases presented here.

\begin{acknowledgments}
We thank B.~Pasquiou for careful reading of the manuscript. We gratefully acknowledge support from the Austrian Ministry of Science and Research (BMWF) and the Austrian Science Fund (FWF) through a START grant under Project No.~Y507-N20. As member of the project iSense, we also acknowledge the financial support of the Future and Emerging Technologies (FET) programme within the Seventh Framework Programme for Research of the European Commission, under FET-Open grant No.~250072.
\end{acknowledgments}

\end{document}